%% file: arxiv.tex
\documentclass[10pt,a4paper]{article}
\input{preamble}

\begin{document}
\include{main_body}
\include{supplement_body}
\end{document}

%% file: preamble.tex
\usepackage[margin=22mm]{geometry}
\usepackage{fontspec}
\usepackage{amsmath,amssymb,graphicx,booktabs,longtable,array,calc}
\usepackage[numbers,sort&compress]{natbib}
\usepackage{chapterbib}
\usepackage[colorlinks=true,allcolors=blue]{hyperref}
\usepackage{caption,fancyhdr}
\makeatletter
\renewcommand{\maketitle}{\begin{center}{\Large\bfseries \@title\par}\ifx\@author\@empty\else\vspace{0.7em}{\normalsize\@author\par}\fi\end{center}\vspace{0.5em}}
\makeatother

%% file: main_body.tex
\title{A Topological Spin on Racetrack Memory: Stability and Transport of 360° Domain Walls in Magnetic Nanotubes}
\author{Benjamin Zingsem and Rafal E. Dunin-Borkowski\\[0.4em]{\small Ernst-Ruska Centre for Microscopy and Spectroscopy with Electrons,\\Forschungszentrum Jülich, 52428 Jülich, Germany}\\[0.4em]{\small Corresponding author: \href{mailto:b.zingsem@fz-juelich.de}{b.zingsem@fz-juelich.de}}}\date{}
\maketitle
\input{sections/abstract}
\input{sections/introduction}
\input{sections/theory}
\input{sections/fig1}
\input{sections/model_and_approach}
\input{sections/results}

\input{sections/summary}
\input{sections/end_statements}

\clearpage
\bibliography{references}

%% file: sections/abstract.tex
\begin{abstract}
Magnetic racetrack memory requires localized carriers that combine rapid motion, resistance to perturbations and a detectable signal. Here we show how a 360\(^\circ\) domain wall in a hollow ferromagnetic tube combines these properties within a single continuous magnetic shell. Two successive magnetization reversals form a bound object whose surrounding domains point in the same direction. Analytical energetics and micromagnetic simulations establish its finite separation, recovery from tested perturbations and current-driven transport without microscopic Dzyaloshinskii--Moriya interaction or a maintaining field. The texture's winding resists unwinding while radial rotation carries an energy cost, providing a specific control over stability. Relative rotation of the constituent spins accounts quantitatively for additional displacement during current onset. The wall supports motion at hundreds of metres per second, with transient velocities approaching one kilometre per second at stronger drive. A prepared seven-bit pattern retains its carriers and local readout contrast after pulsed transport and recovery. Curved-track and differential-current tests identify radial stiffness and drive uniformity as controls for intact motion. These results establish a physical basis for fast, locally distinguishable tubular magnetic bits and identify a continuous switching pathway that can be targeted by tailored excitation.
\end{abstract}

%% file: sections/introduction.tex
\section{Introduction}\label{sec:introduction}
A magnetic racetrack stores information in a pattern that moves past stationary writing and reading elements. Current-controlled domain-wall shift registers have demonstrated the essential principle: translate the magnetization pattern instead of the recording medium \citep{Parkin2008Racetrack,Hayashi2008ShiftRegister}. A useful magnetic carrier must remain distinct during motion, tolerate perturbations and produce a local signal. Meeting these requirements together is more demanding than achieving a large velocity or stabilizing a texture in isolation.

The geometry of the track helps determine which limitations arise. Conventional domain walls in ferromagnetic strips can undergo Walker breakdown, in which changes of the internal spin configuration interrupt steady motion. Localized skyrmions provide a different carrier, but their current-driven transverse deflection---the skyrmion Hall effect---can bring them to a track edge, where they may be lost \citep{Fert2013SkyrmionsOnTrack,Jiang2017SkyrmionHall,Litzius2017SkyrmionHall,Zhang2016BilayerSkyrmions}. Antiferromagnetically coupled multilayers offer an effective solution: compensated skyrmions have been driven at up to 900 m/s with a vanishing Hall deflection \citep{Pham2024FastSAF}. Such results show the value of compensation while motivating alternative ways of combining localized encoding and controlled motion.

Hollow magnetic tubes change the boundary geometry itself. Their circumference has no lateral edge, and a circumferential spin texture need not pass through a filled magnetic axis. Previous work on individual nanotube walls found suppressed Walker breakdown and access to a regime where Cherenkov-like spin-wave emission changes the wall dynamics; current-driven calculations also predict kilometre-per-second motion \citep{Yan2011FastNanotubes,Hertel2016Ultrafast,Otalora2012CurrentPulses}. Tubular geometry therefore makes a high-speed regime accessible that lower-drive structural instabilities can pre-empt in conventional strips. The associated magnonic scale is set by the material, geometry and available spin-wave modes.

A 360\(^\circ\) domain wall (DW) combines this tubular setting with localized encoding. It comprises two successive 180\(^\circ\) reversals: magnetization reverses through the first constituent and returns to its original direction through the second. A bit can thus be represented by the presence or absence of one composite without changing the magnetization of the distant domains. Planar 360\(^\circ\) walls already exhibit binding, manipulation and current-driven motion, and zero-total-charge skyrmioniums also provide localized carriers \citep{Muratov2008Theory360,Zhang2016_360DW,Su2017Enhanced360,Gartside2016Injection,Goebel2019SkyrmioniumRacetrack}. The proposed tubular 360\(^\circ\) carrier \citep{Zingsem2020Patent} brings together a closed circumference and a bound full-turn texture in one continuous ferromagnetic shell. Geometry-dependent binding of domain-wall pairs has also been investigated in CrSBr nanotubes by Martínez-Carracedo et al. \citep{MartinezCarracedo2026GeometricalConfinement}.

Here we establish the energetic and dynamical basis of that combination. Nonlocal attraction and short-range repulsion bind the constituents; winding and the cost of radial rotation give resistance to interior unwinding. The relaxed object recovers from the tested perturbations, and its internal spin rotations quantitatively explain its current-on motion. We then establish transport and recovery of a prepared information pattern and calculate its local magnetic contrast. Curved-track and nonuniform-current tests identify which deformations limit transport and which energy terms suppress them. The resulting picture connects the carrier's demonstrated capabilities to physical controls over their operating range, without requiring a microscopic chiral interaction or a field that maintains the bit.

%% file: sections/theory.tex
\section{Texture and stabilization}\label{sec:theory}
Figure~\ref{fig:texture} shows the magnetic texture of one tubular 360\(^\circ\) DW. Away from the wall, the spins point along the tube. Through the two 180\(^\circ\) constituents, a same-handed texture continues rotating in the same sense: axial, circumferential, reversed axial, opposite circumferential, and finally axial again. The two circumferential cores consequently point in opposite directions. In contrast, a retracing texture would reverse the rotation sense in its second constituent: its circumferential cores would point in the same direction, and the axial spin trajectory would retrace rather than complete a full turn. This distinction changes which deformations can connect the two walls to the uniform background; the resulting binding or annihilation also depends on their energetic interactions.

Let \(s\) be distance along a straight tube and \(\phi\) the angle around its cross-section. The local unit vectors \(\mathbf e_s\), \(\mathbf e_\phi\) and \(\mathbf e_r\) point along the axis, around the circumference and radially outward. Write the unit magnetization as \(\mathbf m=m_s\mathbf e_s+m_\phi\mathbf e_\phi+m_r\mathbf e_r\). At fixed \(\phi\), the projected vector \((m_s,m_\phi)\) traces the spin rotation in the local tangent plane. With \(\psi=m_s+i m_\phi\), its winding is
\begin{equation}
W=\frac{1}{2\pi}\oint \mathrm{d}\!\left(\arg\psi\right),\qquad |\psi|>0.
\label{eq:projected-winding}
\end{equation}
The path closes for periodic boundary conditions or fixed equal magnetization far from the wall. A same-handed full turn has \(W=\pm1\), whereas a retracing pair has \(W=0\).

This winding gives a direct physical picture of stabilization. Rotations that remain in the tangent plane cannot remove the full turn. Small excursions out of that plane also preserve it: changing the winding requires a point where both tangent components vanish and the spin points directly inward or outward. Magnetostatics and exchange make this radial excursion costly in the hollow shell, and an additional radial anisotropy can increase that cost. Anisotropy and exchange therefore provide direct material controls over this obstacle; tube radius and shell thickness also change the exchange and magnetostatic costs. The texture therefore combines a topological restriction on low-energy deformations with an adjustable energetic obstacle to leaving that class. The simulations retain all three spin components, allowing this stabilization to be tested without enforcing exact tangency.

The winding is distinct from the surface topological charge \(Q\), or skyrmion number, which describes how the complete two-dimensional texture covers the sphere of spin directions. For the outward-oriented cylindrical surface, an ideal rotationally symmetric 180\(^\circ\) reversal from \(+\mathbf e_s\) to \(-\mathbf e_s\) carries \(Q=-1\), and the reverse carries \(Q=+1\). The sign records the orientation of the spin-sphere covering; the two constituents have opposite sectors and zero total charge \citep{LeeKim2021Orbital}. Their cancellation leaves the axial winding intact. The zero total charge also permits a smooth route between the full-turn state and a uniformly magnetized tube when radial directions are allowed. Supplementary Section~\ref{sec:S1} gives the surface-charge integral and an explicit finite-energy construction. The physically important question is then how strongly the allowed radial route is opposed, and how selectively an excitation can couple to it.

%% file: sections/fig1.tex
\begin{figure}[htbp]
\centering\includegraphics[width=\linewidth]{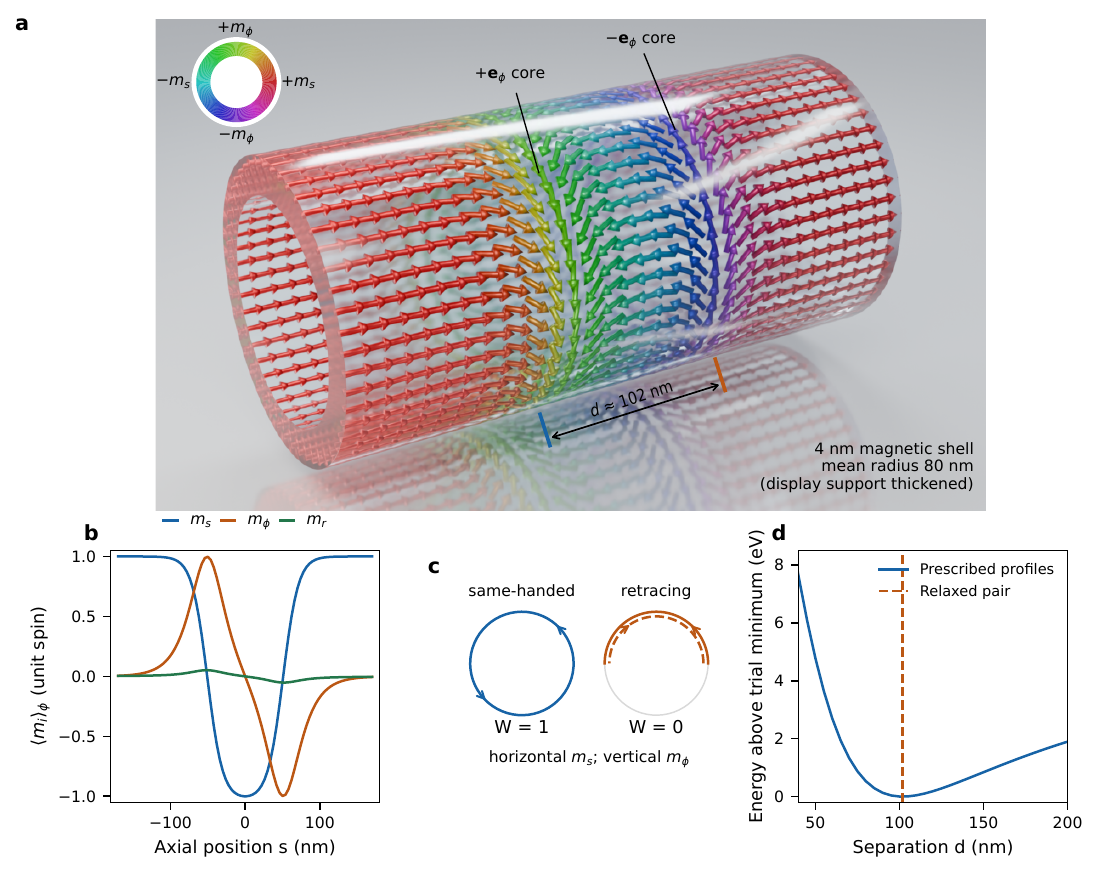}
\caption{\textbf{A bound 360° domain wall in a hollow magnetic tube.}
\textbf{a}, Same-handed 360° domain wall (DW): two successive 180° reversals return to the same axial background. The local unit vectors $\mathbf e_s$, $\mathbf e_\phi$ and $\mathbf e_r$ point along the tube, around its circumference and outward; $m_s,m_\phi,m_r$ are the corresponding unit-spin components. Hue follows the displayed tangent-angle wheel. The transparent support is thickened for visibility; the reference magnetic shell is 4 nm thick. \textbf{b}, Azimuthal means of the saved relaxed mid-shell components, all on the full $[-1,1]$ scale. \textbf{c}, Directions traced by the magnetization while moving along an axial line at fixed $\phi$: projected winding $W=(2\pi)^{-1}\oint \mathrm{d}\!\left[\arg(m_s+i m_\phi)\right]$ is one for the full turn and zero for a retracing pair. The return stroke is offset slightly for visibility. Both full cylindrical textures have zero total surface charge $Q$; the plotted spin-space curve is not the two-dimensional charge integral. \textbf{d}, Finite-thickness prescribed-profile energy above its own minimum, with physical separation in nanometres, and the independently relaxed pair separation. The single-wall inverse-slope parameter of the trial profile is $\Delta_0=20$ nm. Panels a and c are explanatory schematics; b and d use saved numerical data. The straight reference is axially periodic, at zero current and without a maintaining field.}
\label{fig:texture}
\end{figure}

%% file: sections/model_and_approach.tex
\section{Model and approach}\label{sec:model}\label{sec:methods}\label{sec:simulations}
We combine analytical wall profiles, nonlocal magnetostatic calculations and deterministic micromagnetic simulations with MuMax3 \citep{Vansteenkiste2014MuMax}. The main straight-tube model has mean radius 80 nm and a 4 nm magnetic shell. Periodic boundary conditions along the axis represent an extended track without physical end faces. We use exchange stiffness \(A=13\) pJ/m, saturation magnetization \(M_s=0.322\) MA/m and a circumferential hard-axis coefficient \(K_\phi=30.5\) kJ/m\(^3\), giving an exchange length of 14.1 nm. The positive energy density \(K_\phi m_\phi^2\) penalizes circumferential magnetization; the demagnetizing field is computed nonlocally. No microscopic Dzyaloshinskii--Moriya interaction or maintaining external field is included. Thermal noise is omitted and the material parameters are held fixed. The complete energy functional, exact inputs and implementation are given in the \hyperref[sec:extended-methods]{Supplementary Extended Methods}.

Same-handed and retracing spin configurations are initialized separately and allowed to relax without constraining tangency, helicity, symmetry or wall separation. Distinct starting separations and signed perturbations test the relaxed state's restoring response. Escape calculations identify the configurations required to remove its winding. Current-driven trajectories solve the Landau--Lifshitz--Gilbert equation with spin-transfer torque,
\begin{equation}
\partial_t\mathbf m=-\gamma\mathbf m\times\mathbf B_{\rm eff}
+\alpha\mathbf m\times\partial_t\mathbf m
-(\mathbf u\cdot\nabla)\mathbf m
+\beta\mathbf m\times[(\mathbf u\cdot\nabla)\mathbf m].
\label{eq:main-LLG}
\end{equation}
Here \(\gamma>0\) is the gyromagnetic-ratio magnitude, \(\mathbf B_{\rm eff}\) is the energy-derived effective field, \(\alpha=0.1\) is Gilbert damping, and \(\beta=0.05\) measures nonadiabatic spin transfer. The signed spin-advection velocity is \(\mathbf u=-P\mu_B\mathbf J/[eM_s(1+\beta^2)]\), with current polarization \(P=0.4\), conventional current density \(\mathbf J\), Bohr magneton \(\mu_B\) and positive elementary charge \(e\). The equal-coefficient sign/advection control instead uses \(\alpha=\beta=0.1\) (Supplementary Section~\ref{sec:S5}). After current removal, \(\mathbf u=0\) and the same equation governs precession and damping; recovery is physical evolution rather than energy minimization.

We also investigate whether the object can negotiate a curved track. Complete tori remove finite-end effects while introducing different path lengths and current densities on the inside and outside of a bend. To separate those effects, we apply controlled circumferential variations of axial current to straight tubes. These curvature and differential-drive calculations use a separately prepared 14.1 nm shell and a 3 nm cubic mesh. The information-pattern demonstration retains the 4 nm shell with a coarser transverse mesh. Its four prepared textures occupy the seven-bit word $1100101$ plus an empty guard slot, at 384 nm pitch; their physical forward, current-off and reverse phases are followed separately. The \hyperref[sec:extended-methods]{Supplementary Extended Methods} provide the geometry and discretization of each comparison; Supplementary Section~\ref{sec:S6} shows the actual current fields. Only specified straight-track controls include the current's Oersted field; the torus and differential-drive calculations retain full magnetic demagnetization with torque-only driving.

%% file: sections/results.tex
\section{Results and Discussion}\label{sec:results}
\subsection{A bound object with a restoring response}\label{binding-and-radial-escape}
The two constituent 180° walls form a stable composite because their interactions compete. At large separation, the head-to-head and tail-to-tail walls carry opposite effective magnetostatic charges and attract. At short separation, their same-handed rotation profiles repel as they overlap. The balance selects an internal separation, so the constituents remain bound rather than merging or drifting apart. In the stated straight-tube model, the nonlocal energy of prescribed two-wall profiles has a minimum near 103 nm; full-vector micromagnetic relaxation gives approximately 102 nm (Fig.~\ref{fig:texture}). The separation is a property of this geometry and energy balance.

The relaxed state has a measurable restoring response. Three different initial separations approach the same bound branch, and the ten tested signed perturbations return toward the unperturbed reference (Fig.~\ref{fig:escape}a). These tests probe radial tilt, core-spin rotation, circumferential variation, through-thickness variation and combined distortion; their amplitudes and representative returned configurations are shown in Supplementary Section~\ref{sec:S2}. A retracing control instead relaxes to the surrounding domain. This comparison ties the distinction in spin rotation to different physical outcomes. Selected static refinement changes the bound separation by about 0.1\% and the excess energy by about 1.4\%, while preserving the principal structure.

The energetic comparison also explains why the surrounding magnetic orientation matters. We also considered textures whose distant domains follow \(\pm\mathbf e_\phi\), circulating around the tube instead of pointing along its axis. In the axial background studied above, most of the tube is magnetized along its length and each constituent has a leading magnetic pole, producing long-range attraction. The circumferential-background comparison has different charge distributions and need not bind through the same mechanism. The contrasting analytical and simulated cases in Supplementary Sections~\ref{sec:S1} and~\ref{sec:S2} identify the axial-background texture as the bound branch studied here. They also show why neither the total charge nor winding alone determines the equilibrium separation.

\subsection{Fast transport with a quantitative internal-motion contribution}\label{fast-motion-and-the-internal-response}
A bound pair also has a collective dynamical structure. Both constituent walls can translate and rotate their core spins relative to the cylindrical surface. We describe the latter by helicities \(\eta_1\) and \(\eta_2\). At the cores of the displayed same-handed wall, the transverse spin direction is
\begin{equation}
(m_\phi,m_r)\propto \sigma_i(\cos\eta_i,\sin\eta_i),
\qquad \sigma_1=+1,\quad\sigma_2=-1.
\label{eq:core-helicity}
\end{equation}
Both helicities are zero for their reference tangent orientations. This rotation changes the texture relative to the tube surface; rotating the entire ideal axisymmetric spin field together with space leaves it unchanged. The numerical angle estimator is specified in Supplementary Section~\ref{sec:S4}.

The opposite constituent surface topological-charge sectors cancel the leading coupling of pair translation to their common spin rotation. A coupling to the difference \(\delta\eta=\eta_2-\eta_1\) remains. For the constituent axial positions \(X_1,X_2\) and their mean \(C=(X_1+X_2)/2\), projection of spin-transfer dynamics onto position, separation and the two internal angles gives, in the absence of a conservative axial force,
\begin{equation}
\dot C\simeq\frac{\beta}{\alpha}u-c_\eta\dot{\delta\eta}.
\label{eq:pair-transport}
\end{equation}
Here \(u\) is the signed spin-advection velocity set by the current, and \(c_\eta\simeq91.4\) nm/rad is calculated from the relaxed vector field. The first term gives motion with fixed internal angles; the second is displacement associated with their relative rotation. The short reduction from the projected force balance and the full tensors are given in Supplementary Section~\ref{sec:S4} \citep{Thiaville2005STT,Tretiakov2008Collective}.

Periodic transport directly tests this relation (Fig.~\ref{fig:transport}). At \(J_s=10^{12}\) A/m\(^2\), the independently calculated coefficient combined with the measured angle change accounts for approximately 96\% of the additional displacement over the first 0.5 ns for spin torque alone, and 95\% when the ideal shell-current self-field is included. As the relative angle settles, the speed decreases toward 36.0 m/s, close to the fixed-angle prediction of 35.9 m/s. The current is switched on abruptly from a prepared stationary state. Its initial spin-transfer response is predominantly advection at speed \(|u|\), whereas settled internal angles leave \(|\beta u/\alpha|\simeq |u|/2\). The relative-angle motion supplies the initial excess translation and then decays. Because Eq.~\eqref{eq:main-LLG} is first order in time, an ideal current step can change the wall velocity immediately without a Newtonian acceleration stage. The first plotted velocity averages the first 5 ps; the self-field additionally produces a small resolved early overshoot (Supplementary Section~\ref{sec:S4}). The bound composite retains the usual settled spin-transfer mobility while its transient displacement has an independently quantified internal origin.

Increasing the current produces fast motion over the sampled intervals (Fig.~\ref{fig:transport}). At \(2\), \(4\) and \(8\times10^{12}\) A/m\(^2\), the speeds fitted over 0.04--0.30 ns after current-on are approximately 117, 233 and 473 m/s; the wall remains intact through the respective 1.2, 0.9 and 0.7 ns observations. At \(16\times10^{12}\) A/m\(^2\), the same fitting procedure gives approximately 1.0 km/s before subsequent annihilation, identifying this highest speed as transient. Previous nanotube studies established access to very high wall speeds with suppressed conventional Walker breakdown and a Cherenkov-like spin-wave-radiation regime \citep{Yan2011FastNanotubes,Hertel2016Ultrafast}. The present current series measures the composite's transient response; its own radiation onset remains to be determined. All fitting windows and endpoint observations appear in Supplementary Section~\ref{sec:S5}.

Current-induced compression means a decrease of the constituent separation \(d=X_2-X_1\): the reversed axial domain narrows. Spin transfer excites the internal angles and changes their coupled force balance with separation; an azimuthal Oersted field also acts oppositely on the two circumferential cores. In the matched \(8\times10^{12}\) A/m\(^2\) trajectories, the modeled Oersted field enhances both displacement and compression. Thus the same internal freedom that contributes to motion must remain energetically confined during sustained driving. The finite-ended 180\(^\circ\)/360\(^\circ\) control in Supplementary Section~\ref{sec:S5} gives similar displacements, whereas the periodic trajectories above provide the cleaner test of the internal-motion relation.

\input{sections/fig3}
\subsection{A transported pattern with local readout contrast}\label{transported-information-and-local-detection}
The competition between long-range constituent attraction and short-range overlap repulsion established in Section~\ref{binding-and-radial-escape} gives each 360° wall a restoring internal separation. The cancellation of its leading magnetic poles also changes the interaction between well-separated composites (Supplementary Sections~\ref{sec:S2} and~\ref{sec:S9}). We next test whether these bound carriers can retain an information pattern while moving together. Four 360\(^\circ\) DWs occupy the `1' positions in the seven-bit binary representation of decimal 101, \(1100101\), with one additional empty guard slot in the periodic track. This pattern includes adjacent occupied slots and occupied slots separated by one or two zeros. The nominal pitch is 384 nm. A high-current pulse moves every texture by more than one pitch while preserving their order over the observation; a lower-current pulse provides the clearer recovery test (Fig.~\ref{fig:word}).

After 2 ns at \(-4\times10^{12}\) A/m\(^2\), all four textures have moved approximately 348--359 nm. With the current removed for 0.8 ns, their internal profiles recover and all four carriers remain distinct and ordered. Recoil leaves net displacements of 288--306 nm; the largest deviation from common displacement is approximately 10 nm, less than 3\% of one slot pitch. This is a demonstration of pattern transport and recovery, with measurable positional relaxation. A subsequent 1 ns pulse of opposite sign, \(+4\times10^{12}\) A/m\(^2\), translates all four textures back by a further 205--216 nm while preserving their identity (Supplementary Movie~\ref{movie:train}). The stronger-drive comparison in Supplementary Section~\ref{sec:S7} establishes why pulse duration and recovery should be considered together.

Our magnetization fields also predict a substantial local magnetic signal. For the single 4 nm-shell pair, the calculated stray-field contrast reaches approximately 10 mT at a point 40 nm outside the tube (Supplementary Section~\ref{sec:S9} and Fig.~\ref{fig:point-readout}). A finite detector footprint retains millitesla-scale axial contrast at the investigated separations (Fig.~\ref{fig:readout}b). Opposite wall poles cancel the leading isolated charge contribution at long range, while their local fields remain accessible. The isolated-pair diagnostic therefore combines local magnetic contrast with a dipolar far field, which decays faster than the leading pole field of an isolated charged wall. The pole cancellation follows analytically from the paired reversals, while the saved-field calculation demonstrates their accessible near-field contrast. This intrinsic combination of reduced remote field, local readability and fast transport is useful for a localized carrier; quantitative crosstalk at a chosen pitch still depends on the full field.

The recovered pattern has a second, complementary signal in passive magnetoresistance estimates. Anisotropic magnetoresistance weights the squared axial magnetization and resolves the rotation in the wall cores; an axial-reference orientation contrast resolves the intervening reversed regions \citep{McGuire1975AMR,CorteLeon2014AMR}. Both retain distinct spatial features in the actual recovered state. These are our calculated field and normalized resistive contrasts; voltage, noise and detector back-action require a specified reader. Their coexistence with binding and current-driven recovery establishes the physical ingredients for a locally addressable tubular carrier.
\input{sections/fig4}
\input{sections/fig5}
\subsection{Radial rotation provides a tunable switching coordinate}\label{sec:radial-control}
The bound wall resists interior unwinding until a local region rotates sufficiently far toward the surface normal. Along the lowest-energy route identified, a confined patch of spins, localized both along the tube and around its circumference, makes this excursion (Fig.~\ref{fig:escape}). The transition configuration lies about 2.4 eV above the bound state in the 4 nm-shell model. Exchange and magnetostatic costs of about 2.7 and 2.5 eV outweigh a 2.8 eV decrease in circumferential anisotropy energy. A coherent radial rotation of the whole ring reaches a sampled cost of approximately 7.3 eV, although that comparison is not saddle-converged. The localized route therefore explains how the surrounding texture can remain largely intact while a small region crosses the energetic obstacle.

The tangent projection is still nonzero at the identified energy maximum. Beyond that maximum, successive configurations toward the uniform state have decreasing energy; winding changes on this descending part of the path when the local magnetization reaches a radial direction. This sequence connects the topology to a concrete magnetic phase slip \citep{Kim2016PhaseSlips}. The full magnetization remains smooth at the resolution of the sampled pathway. The 2.4 eV value is an approximate barrier for this reference model; residual-gradient and spatial/path-convergence qualifications are given in Supplementary Section~\ref{sec:S3}.

This escape geometry is also constructive information for control. A sufficiently strong localized excitation could nucleate a radial phase slip, after which the remaining wall could unwind along the descending path without coherent reversal around the entire circumference. This requires crossing the local energetic obstacle; the tested small radial-tilt perturbations instead relax back. The identified pathway and its response to differential spin torque provide a strong mechanistic indication that tailored spatial or temporal excitation could switch the texture efficiently. Shaped pulses or resonant excitation of a coupled internal mode are natural candidates: mode-selective switching has precedent in magnetic vortices \citep{VanWaeyenberge2006VortexReversal,Kammerer2011SpinWaveReversal}. Here it is a motivated control prospect; the excitation spectrum, source waveform and dissipated energy of such a protocol have not been determined. The simple local-field pulse tests, including their source-specific outcomes, are reported in Supplementary Section~\ref{sec:S8}.

\input{sections/fig2}
\subsection{Robust transport through nonuniform and curved tracks}\label{uneven-drive-and-a-stabilizing-intervention}
A useful carrier must accommodate more than a perfectly uniform track. The 4 nm-shell wall traverses both tested local increases and decreases of circumferential anisotropy, with Gaussian perturbations of amplitude 25\% at a current of \(4\times10^{12}\) A/m\(^2\). Separate 20\% increases or decreases of exchange or circumferential anisotropy also retain the bound pair over the tested driven intervals. These controlled perturbations show that its existence and transport are not confined to one exact numerical parameter choice. Supplementary Section~\ref{sec:S7} shows their profiles and measured responses.

Curved routing introduces a different challenge. For a wall to progress coherently around a bend, its outer segments must cover more distance and therefore move faster than its inner segments. The current of a homogeneous toroidal conductor is instead strongest inside. Current crowding reinforces the geometric mismatch, and exchange, anisotropy and magnetostatics must keep the circumferential segments moving together. Three-dimensional current maps and a straight-tube control with circumferentially varying axial current make these effects explicit in Supplementary Section~\ref{sec:S6}.

The 14.1 nm-shell torus supports intact transport under the physical current profile. At a mean current magnitude of \(10^{12}\) A/m\(^2\) and major radius 704 nm, the extended trajectory retains the wall through approximately 490 nm of travel over 9.44 ns, corresponding to a 40\(^\circ\) arc (Supplementary Movie~\ref{movie:torus}). A separate 1,056 nm-radius torus retains the wall over its 200 nm observation. These results demonstrate coherent transport across a substantial curved segment. A late radial deformation appears near the end of the extended run; its mechanism and numerical sensitivity remain unresolved (Supplementary Section~\ref{sec:S6}). The matched comparisons give useful design directions: reducing current extends intact travel at fixed radius, and increasing the radius reduces measured shear over the common first 200 nm. A smaller tube-to-bend-radius ratio also reduces current crowding and the inner/outer path-length mismatch; decreasing tube diameter is a prospective way to reduce that ratio, but would also change the bound-state energetics. At twice the current in the smaller torus, replacing the inner-enhanced current by a constant magnitude delays the first projection-screen event, and a prescribed current-density profile compensating the angular-advection mismatch preserves the wall over the tested distance. The different profiles retain the same magnetic geometry and reveal how differential forcing contributes to the operating boundary. Related current-crowding and unequal-path-length effects are known in other curved tracks \citep{Garg2017Curvature,OsunaRuiz2021Curved,Nunes2026BentTubes}.

We isolate differential forcing in a straight tube by imposing an axial current that varies by 25\% around the circumference. The unrolled current, initial magnetization and a radial-escape state are shown together in Supplementary Fig.~\ref{fig:differential-maps}. Under this nonuniform drive, an added radial hard-axis term \(K_r m_r^2\), with \(K_r=16.25\) kJ/m\(^3\), preserves the wall through approximately 200 nm. A 20\% increase of exchange alone delays the event. Uniform-current transport in the same 14.1 nm straight shell is visualized in Supplementary Movie~\ref{movie:straight}. The radial term therefore stabilizes the tested straight-tube differential-drive motion. Its extension to a torus is a design prospect: radial confinement does not remove the unequal inner and outer path lengths, and a toroidal test with this added anisotropy has not been performed. Together, radial confinement, gentler curvature and a more uniform angular drive identify a route toward extending coherent transport.

%% file: sections/fig3.tex
\begin{figure}[htbp]
\centering\includegraphics[width=\linewidth]{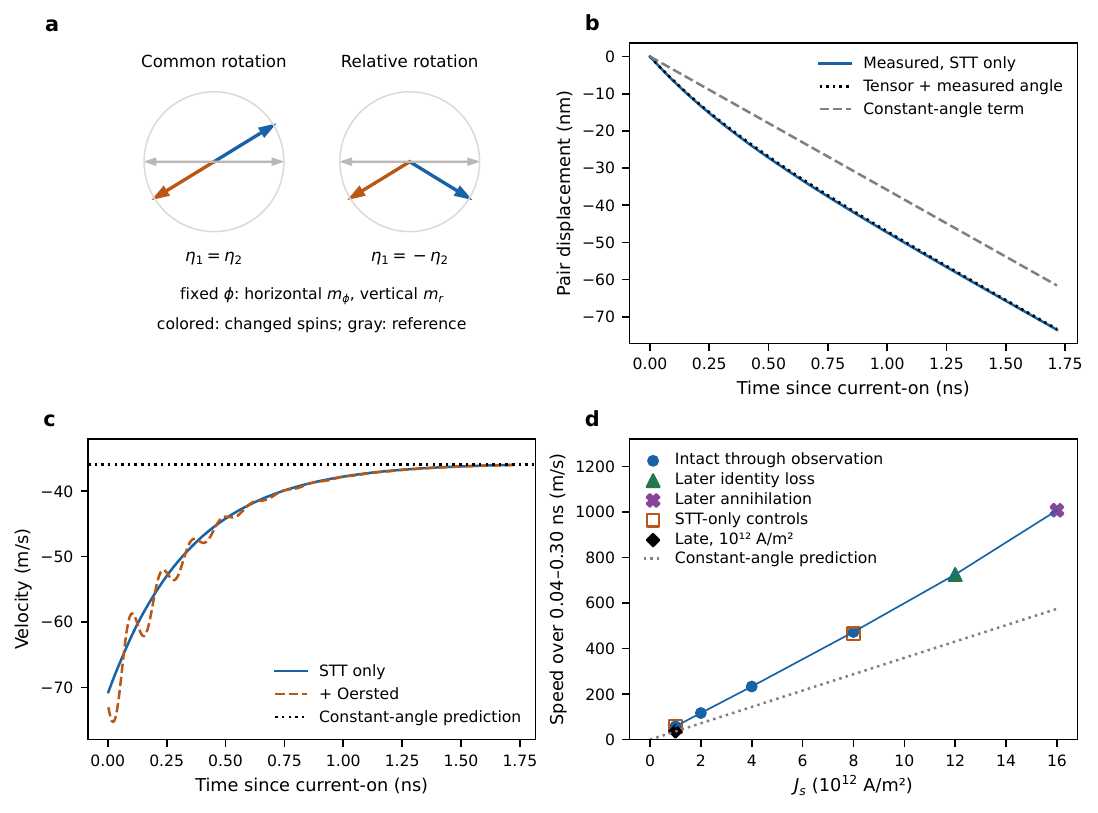}
\caption{\textbf{Relative spin rotation contributes to translation.}
\textbf{a}, Schematic constituent spins at fixed $\phi$. Gray arrows show reference core directions; blue and orange arrows show changed spins in the $(m_\phi,m_r)$ plane. Common and opposite helicity changes are rotations relative to the local frame, not rigid rotation of the whole axisymmetric field. \textbf{b}, Measured periodic-pair displacement at $J_s=10^{12}$ A/m$^2$, compared with the constant-angle contribution and Eq.~\eqref{eq:pair-transport} using the independently integrated $c_\eta$ and measured relative-angle history. The latter is an input to the balance, not a separately predicted trajectory. \textbf{c}, The prepared state is at rest before the current step at $t=0$. Finite-difference velocities begin with the 0--5 ps mean, plotted at its midpoint. Initial motion is approximately the advection velocity, and internal-angle settling reduces it toward $-35.9$ m/s; final-window fits give approximately $-36.0$ m/s. \textbf{d}, Speed versus current fitted over 0.04--0.30 ns after switch-on for the same 4 nm periodic reference, using only valid positions. Circles are intact through the recorded observations (2/4/8 × $10^{12}$ A/m$^2$: 1.2/0.9/0.7 ns); the triangle marks a later projection-screen event at $12\times10^{12}$ A/m$^2$ (0.509 ns), whose terminal fate remains unresolved; the cross marks $16\times10^{12}$ A/m$^2$ loss (0.376 ns) followed by annihilation. Open squares denote torque-only controls and the diamond the separately measured late value at $10^{12}$ A/m$^2$. The dotted line is $|v|=(\beta/\alpha)|u|$ from Eq.~\eqref{eq:pair-transport}, derived in Supplementary Section~\ref{sec:S4} from spin-transfer dynamics \citep{Thiaville2005STT,Tretiakov2008Collective}. Panel d plots speed magnitudes. The projection screen requires the minimum tangent norm to exceed 0.05 in both occupied-cell and reconstructed-surface diagnostics (\hyperref[sec:projection-observables]{Supplementary Extended Methods}). All physical dynamics use $\alpha=0.1$, $\beta=0.05$ and $P=0.4$, defined with Eq.~\eqref{eq:main-LLG}.}
\label{fig:transport}
\end{figure}

%% file: sections/fig4.tex
\begin{figure}[htbp]
\centering\includegraphics[width=\linewidth]{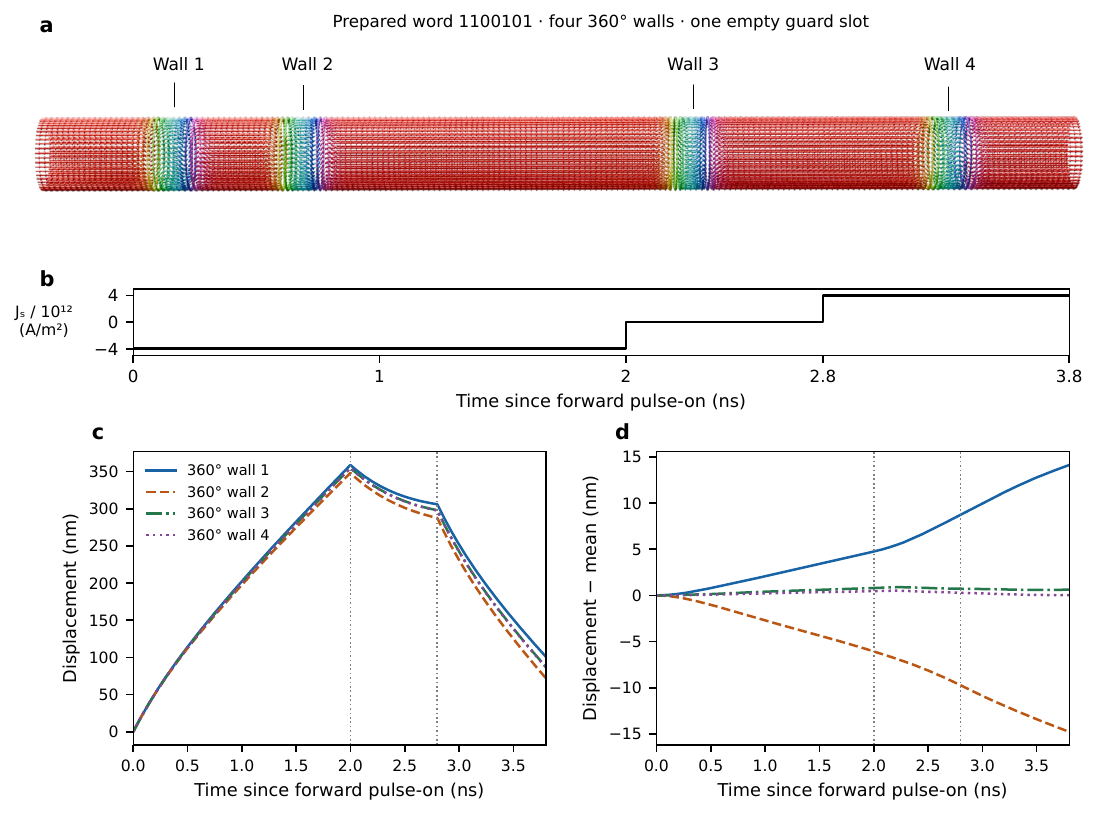}
\caption{\textbf{A prepared information pattern moves forward, recovers and reverses.}
\textbf{a}, Actual initial magnetization of four 360° DWs encoding $1100101_2=101_{10}$ in seven data slots, with an eighth empty guard slot between periodic copies. The track has no physical end faces at the displayed limits. Spin hue follows Fig.~\ref{fig:texture}. Labels identify the four 360° walls occupying the `1' slots. The nominal pitch is 384 nm. \textbf{b}, Applied conventional-current protocol: $-4\times10^{12}$ A/m$^2$ for 2 ns, zero current for 0.8 ns, then $+4\times10^{12}$ A/m$^2$ for 1 ns. The reverse phase starts from the exact recovered checkpoint, without relaxation or minimization. \textbf{c}, Independently tracked, unwrapped displacements of all four textures. \textbf{d}, Each displacement relative to the contemporaneous four-texture mean, exposing spacing drift. The clock begins at forward pulse-on; dotted lines mark current changes. All four survive through 3.8 ns, including 205--216 nm of reverse-phase travel. These are finite-time motions, not steady mobilities or optimized pulses. The 4 nm shell is discretized at $4\times2\times2$ nm; damping $\alpha=0.1$, nonadiabaticity $\beta=0.05$ and polarization $P=0.4$ remain fixed; the drive is torque-only, without an Oersted field. Supplementary Movie~\ref{movie:train} shows the same sequence.}
\label{fig:word}
\end{figure}

%% file: sections/fig5.tex
\begin{figure}[htbp]
\centering\includegraphics[width=\linewidth]{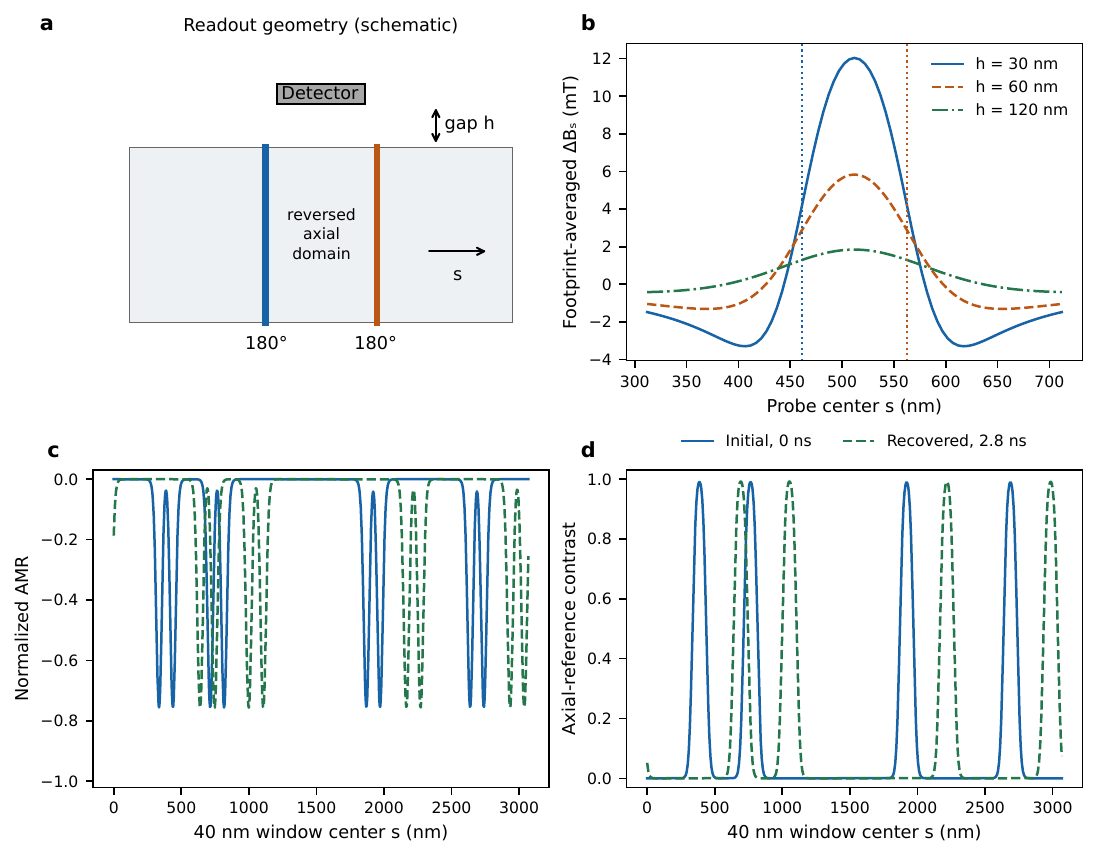}
\caption{\textbf{The simulated carrier provides local magnetic and resistive contrast.}
\textbf{a}, Schematic axial view of the two constituent walls and an external detector. The surface gap is $h$; the footprint is $40\times40$ nm in a plane parallel to the tube axis. \textbf{b}, Our calculated footprint-averaged axial stray-field contrast $\Delta B_s$ from the stationary single-pair 4 nm reference, relative to its no-bit background. The peak magnitudes are 12.03, 5.83 and 1.85 mT at the stated gaps. Blue and orange vertical lines mark the two measured constituent centers, aligned with panel a. The field is reconstructed from the saved micromagnetic magnetization, not an analytical trial wall; the source quadrature and validation are in Supplementary Section~\ref{sec:S9}, with the point-field scan in Supplementary Fig.~\ref{fig:point-readout}. No drive current is applied. These external fields are distinct from the internal effective field. \textbf{c,d}, Passive contrast calculated from the actual initial and recovered train fields of Fig.~\ref{fig:word}. The normalized anisotropic magnetoresistance (AMR) is $\langle m_s^2-m_{s,\mathrm{bg}}^2\rangle$; the axial-reference orientation contrast is $\langle(m_{s,\mathrm{bg}}-m_s)/2\rangle$. Brackets mean equal cross-sectional weighting and a 40 nm periodic axial window. Window centers are plotted without re-registering individual bits. The 2.8 ns field follows 2 ns of forward drive and 0.8 ns at zero current. AMR resolves rotating cores; orientation contrast also resolves reversed domains. The calculated quantities and averaging operations are defined in Supplementary Section~\ref{sec:S9}.}
\label{fig:readout}
\end{figure}

%% file: sections/fig2.tex
\begin{figure}[htbp]
\centering\includegraphics[width=\linewidth]{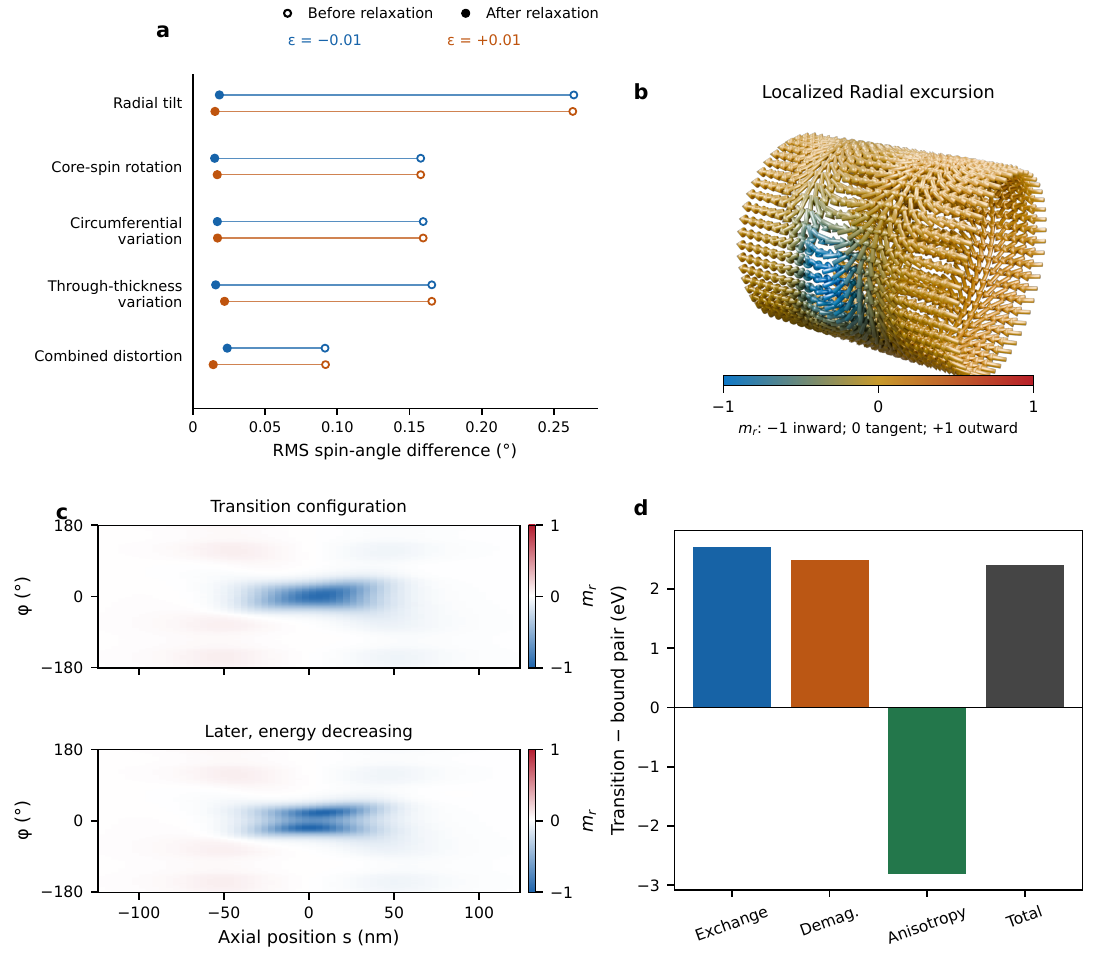}
\caption{\textbf{A restoring bound branch with a localized radial escape coordinate.}
\textbf{a}, RMS spin-angle difference from one unperturbed relaxed reference, before (open symbols) and after (filled symbols) free relaxation. Both signs of five small distortions are tested: radial tilt, core-spin rotation, circumferential variation, through-thickness variation and combined distortion. The maximum imposed spin-vector change is approximately 0.01 ($\epsilon=\pm0.01$). Angles are measured over all occupied cells after removing only common axial translation. Initial differences of 0.092°--0.264° decrease to 0.014°--0.024°. The residual distortions remain resolved. These are ten deterministic returns, not a statistical ensemble. No perturbing field is maintained. The reference, solver controls and additional energy, separation and winding checks are defined in Supplementary Section~\ref{sec:S2}; numerical relaxation is not physical time. \textbf{b}, Three-dimensional magnetization at the identified transition configuration of the 4 nm reference. Arrow directions show the sampled spins, and colour indicates the radial component $m_r$. The inward nucleus has $m_r<0$. \textbf{c}, The middle surface unrolled into axial position $s$ and angle $\phi$, at the transition and later on the descending part of the energy path toward the uniform state. Radial magnetization uses a blue--white--red scale: $-1$ inward, zero tangent, $+1$ outward. These are ordered energy-path states, not equal-time frames. \textbf{d}, Energy-component changes from the bound pair to the transition, with approximate total cost 2.4 eV. Winding loss follows the energy maximum (Supplementary Section~\ref{sec:S3}). All panels use the same unbiased 4 nm model; no current maintains these states.}
\label{fig:escape}
\end{figure}

%% file: sections/summary.tex
\section{Discussion and outlook}\label{sec:discussion}
The tubular 360\(^\circ\) DW combines localized encoding, internal binding and selective dynamical compensation within one continuous magnetic shell. The hollow geometry removes the lateral edge-loss route of a planar track; the paired texture leaves the surrounding axial domains unchanged. These features accompany current-driven motion, recovery from tested perturbations and calculated local readout contrast. They are useful in combination even though individual advantages also occur in planar 360\(^\circ\) walls, skyrmioniums and compensated multilayers. The present quantitative contribution is the connection between the constituent interactions, winding, internal rotation and an identified radial control coordinate.

Stabilization requires neither microscopic Dzyaloshinskii--Moriya interaction nor tuning to a magnetic compensation point. The bound state follows from exchange, anisotropy and magnetostatics in a ferromagnetic shell, without requiring access to a chiral skyrmion phase. This provides a route to localized carriers whose material requirements are expressed through the balance of those energy scales and geometry. It motivates investigating broad operating intervals within the ferromagnetic phase. The actual interval will depend on the temperature evolution of the material parameters and escape barrier. Engineered-film skyrmions also operate at room temperature \citep{Woo2016RoomTemperature,Boulle2016RoomTemperature}.

The smooth radial pathway further suggests that retention and selective switching can be addressed through different forms of excitation. A spatially tailored field, deliberate differential torque or resonant coupling to an internal mode could concentrate the driving action on the relevant deformation. The exchange and radial-anisotropy controls provide a concrete starting point for this prospect. Stronger radial confinement in transport sections and selective excitation at addressing sites may consequently be compatible, with source efficiency, switching selectivity and the preparation of new bits left to a dedicated control study.

\section{Summary}\label{sec:summary}
Analytical energetics and full-vector simulations establish a bound tubular 360\(^\circ\) domain wall that recovers from the tested perturbations, moves under spin-polarized current and provides local magnetic contrast. Its winding restricts low-energy unwinding, while radial rotation supplies an energetically adjustable route for changing the state. The constituent spin rotations account quantitatively for transient displacement, and controlled changes of radial stiffness and drive uniformity identify ways to preserve coherent motion. Prepared-pattern recovery and intact curved-track transport demonstrate the relevance of these mechanisms beyond an isolated stationary wall. Together, the results provide a physical basis for developing fast, locally distinguishable magnetic carriers and for tailoring their stability and switching response.

%% file: sections/end_statements.tex
\section*{Acknowledgements}\label{sec:acknowledgements}
This project has received funding from the European Research Council (ERC) under the European Union's Horizon 2020 research and innovation programme (Grant No. 856538, project “3D MAGiC”). The authors acknowledge helpful discussions with Michael Farle, Thomas Feggeler, Irene Iglesias, and Oleg Tretiakov.

\section*{Conflict of interest}\label{sec:conflict-of-interest}
The authors declare no conflict of interest.

\section*{Data availability}\label{sec:data-availability}
The data supporting the findings of this study are available from the corresponding author on reasonable request.

%% file: supplement_body.tex
\setcounter{section}{0}
\setcounter{equation}{0}
\setcounter{figure}{0}
\setcounter{table}{0}
\renewcommand{\theHsection}{S\arabic{section}}
\renewcommand{\theHequation}{S\arabic{equation}}
\renewcommand{\theHfigure}{S\arabic{figure}}
\renewcommand{\theHtable}{S\arabic{table}}
\renewcommand{\theequation}{S\arabic{equation}}
\renewcommand{\thesection}{S\arabic{section}}
\renewcommand{\thefigure}{S\arabic{figure}}
\title{Supplementary Information\\A Topological Spin on Racetrack Memory: Stability and Transport of 360° Domain Walls in Magnetic Nanotubes}
\author{Benjamin Zingsem and Rafal E. Dunin-Borkowski\\[0.4em]{\small Ernst-Ruska Centre for Microscopy and Spectroscopy with Electrons,\\Forschungszentrum Jülich, 52428 Jülich, Germany}\\[0.4em]{\small Corresponding author: \href{mailto:b.zingsem@fz-juelich.de}{b.zingsem@fz-juelich.de}}}\date{}
\maketitle
\input{sections/methods_extended}
\input{sections/S1}
\input{sections/S2}
\input{sections/S3}
\input{sections/S4}
\input{sections/S5}
\input{sections/S6}
\input{sections/S7}
\input{sections/S8}
\input{sections/S9}
\input{sections/S10}
\input{sections/movies}
\clearpage
\input{sections/supporting_figures}
\clearpage
\bibliography{references}

%% file: sections/methods_extended.tex
\section*{Extended Methods}\label{sec:extended-methods}
\addcontentsline{toc}{section}{Extended Methods}
\subsection*{Magnetic energy and model families}
The magnetization is \(\mathbf M=M_s\mathbf m\), with \(|\mathbf m|=1\) and saturation magnetization \(M_s\). In the magnetic region \(\Omega\), the reference energy is

\begin{equation}
\label{eq:energy}
E=\int_\Omega\!\left[A|\nabla\mathbf m|^2+K_\phi(\mathbf m\cdot\mathbf e_\phi)^2-M_s\mathbf B_{\rm ext}\cdot\mathbf m\right]dV
-\frac{M_s}{2}\int_\Omega\mathbf m\cdot\mathbf B_d\,dV.
\end{equation}

Here \(A\) is the exchange stiffness, \(K_\phi>0\) penalizes circumferential magnetization, \(\mathbf e_\phi\) is the local circumferential unit vector, \(\mathbf B_{\rm ext}\) is an applied field and \(\mathbf B_d\) is the full nonlocal demagnetizing field. The parameters are \(A=13\) pJ/m, \(M_s=321{,}637.549\) A/m and \(K_\phi=30{,}468.327\) J/m³. They define an illustrative magnetic model rather than a fit to a particular fabricated material. The exchange length \(\ell_{\rm ex}=\sqrt{2A/(\mu_0M_s^2)}\) is 14.142 nm, where \(\mu_0\) is the vacuum permeability. The reference includes neither microscopic Dzyaloshinskii--Moriya interaction nor a field that maintains the prepared texture.

{\def\LTcaptype{table} 
\begin{longtable}[]{@{}
  >{\raggedright\arraybackslash}p{(\linewidth - 6\tabcolsep) * \real{0.2500}}
  >{\raggedright\arraybackslash}p{(\linewidth - 6\tabcolsep) * \real{0.2500}}
  >{\raggedright\arraybackslash}p{(\linewidth - 6\tabcolsep) * \real{0.2500}}
  >{\raggedright\arraybackslash}p{(\linewidth - 6\tabcolsep) * \real{0.2500}}@{}}
\toprule\noalign{}
\begin{minipage}[b]{\linewidth}\raggedright
Calculation
\end{minipage} & \begin{minipage}[b]{\linewidth}\raggedright
Magnetic geometry
\end{minipage} & \begin{minipage}[b]{\linewidth}\raggedright
Cell dimensions
\end{minipage} & \begin{minipage}[b]{\linewidth}\raggedright
Boundary conditions
\end{minipage} \\
\midrule\noalign{}
\endhead
\bottomrule\noalign{}
\endlastfoot
Reference state, escape and periodic single-pair transport & Inner/outer radii 78/82 nm; length 1,024 nm & 4 × 1 × 1 nm; selected static refinement 2 × 0.5 × 0.5 nm & Axial periodicity; demagnetizing repeat count 4 \\
Finite single-wall comparison and local-loop pulses & Same shell and length & 4 × 1 × 1 nm & Unconstrained magnetic ends; no Cartesian periodicity \\
Seven-bit pattern, represented by four textures & Same shell; length 3,072 nm & 4 × 2 × 2 nm & Axial periodicity; 384 nm slot pitch and one empty guard slot \\
Differential-drive and curvature controls & Inner/outer radii 72.929/87.071 nm & 3 nm cubic cells & Periodic straight tube or complete closed torus, as specified \\
\end{longtable}
}

Figure~\ref{fig:model-geometries} distinguishes the smooth geometry from the Cartesian discretization. The thick straight tube is 1,152 nm long. The tori have major radii 704 and 1,056 nm, with Cartesian grids 540 × 540 × 63 and 768 × 768 × 63. Their magnetic tracks close geometrically; Cartesian periodic boundaries are disabled. Results from different shell thicknesses are not a mesh-refinement series.

The radial-anisotropy intervention adds \(K_r(\mathbf m\cdot\mathbf e_r)^2\), where \(\mathbf e_r\) is the local surface normal and \(K_r=0.25K_d=16.25\) kJ/m³, with \(K_d=\mu_0M_s^2/2\). This is an additional material anisotropy; full demagnetization is retained. Other specified sensitivity tests vary \(A\) or \(K_\phi\) separately.
\subsection*{Micromagnetics, preparation and dynamics}
Calculations use MuMax3 version 3.12, commit \texttt{6e5c98bb\allowbreak{}66735e09\allowbreak{}d35f2814\allowbreak{}5f337755\allowbreak{}c0893e79}, with binary cell occupancy \citep{Vansteenkiste2014MuMax}. The full demagnetizing interaction is retained. Same-handed and retracing initial states are constructed separately. Initialization fields are removed before their source-free stability is assessed. Relaxation does not constrain tangency, axial symmetry, helicity, radial uniformity or constituent separation. Signed perturbations probe radial tilt, core-spin rotation, circumferential variation, through-thickness variation and combined distortion. The selected fine-grid state is additionally checked after near-state minimization and subsequent relaxation; quantitative static-resolution tests appear in Supplementary Section~\ref{sec:S2}.

Physical trajectories use deterministic Landau--Lifshitz--Gilbert dynamics with thermal noise omitted and material parameters held fixed. For uniform axial spin transfer,

\begin{equation}
\label{eq:LLG}
\partial_t\mathbf m=-\gamma\mathbf m\times\mathbf B_{\rm eff}
+\alpha\mathbf m\times\partial_t\mathbf m
-u\partial_s\mathbf m+\beta u\,\mathbf m\times\partial_s\mathbf m.
\end{equation}

The positive magnitude of the gyromagnetic ratio is \(\gamma=1.7595\times10^{11}\) rad/(T s); \(\mathbf B_{\rm eff}\) is the energy-derived effective field in tesla. Unless stated otherwise, Gilbert damping is \(\alpha=0.1\), the nonadiabatic coefficient is \(\beta=0.05\), and current polarization is \(P=0.4\). The pinned solver convention gives

\begin{equation}
\label{eq:advection}
u=-\frac{P\mu_BJ_s}{eM_s(1+\beta^2)},
\end{equation}

where \(u\) is the spin-advection velocity and \(J_s\) is conventional axial current density, \(e>0\) is the elementary charge and \(\mu_B\) is the Bohr magneton. Positive \(J_s\) therefore drives negative-\(s\) advection. For spatially varying vector current, \(u\partial_s\) is replaced by \(\mathbf u\cdot\nabla\) in the corresponding local spin-transfer terms. The current is prescribed; electronic spin accumulation or texture-dependent current redistribution is not solved.

Precessional dynamics use adaptive RK45 integration with \texttt{MaxErr=1e-7}. A physical current-on or current-off trajectory uses the stated dynamic damping and no relaxation or minimization. Pulse-off recovery is therefore real simulated evolution, not an optimized endpoint. Preparation damping is solely a numerical parameter. Omitting stochastic fields isolates the deterministic bound branch, internal response and escape pathway. Thermal lifetimes are not inferred from these trajectories; the conditional activated-escape interpretation is given in Supplementary Section~\ref{sec:S10}.
\subsection*{Current fields and comparisons}
Selected straight runs include the self-field of uniform axial current in an annular magnetic shell,

\begin{equation}
\label{eq:oersted}
B_\phi(r)=\frac{\mu_0J_s(r^2-R_i^2)}{2r},\qquad R_i<r<R_o,
\end{equation}

with no core current. Matching torque-only controls omit this field. The curved and differential-drive calculations retain magnetic demagnetization but omit current self-fields, contacts, Joule heating and return-path fields.

For an ideal homogeneous torus, the tangential current is \(\mathbf J=(C_J/\rho)\mathbf e_\vartheta\), normalized so its cross-sectional mean equals the quoted current density. The major-axis distance is \(\rho\), and \(\mathbf e_\vartheta\) points around the torus. Constant-magnitude and \(J\propto\rho\) profiles are prescribed mechanism controls, not assumed current distributions of the homogeneous conductor. Supplementary Section~\ref{sec:S6} gives the geometry, straight-tube differential-current profile, normalization and wall-line harmonic definition. Source/drive comparisons restart the same accepted state unless a physical parameter is changed; changed-energy models are prepared independently.
\subsection*{Escape paths, positions and topological observables}\label{sec:projection-observables}
The escape study combines manifold-respecting interpolation and energy descent with a climbing-image refinement, using the native energy and field evaluator \citep{Bessarab2015GNEB, E2007String}. Opposite descents, local directional-curvature checks and a low-energy return connection support the transition configuration. Residual-gradient and resolution limits are reported with the energy estimate rather than interpreted as a converged global minimum barrier.

Longitudinal zero crossings identify constituent walls. Their separation is distinguished from distances between the centers of separate 360° bits. Individual bits are followed across periodic seams. Helicity is a weighted spin angle relative to the reference core orientation; the estimator, temporal unwrapping and separate tensor-coordinate partition are defined in Supplementary Section~\ref{sec:S4}. Surface charge is evaluated by oriented solid-angle triangles on inner, middle and outer sampling surfaces \citep{BergLuscher1981Charge}. Projected winding is evaluated only where the tangent projection and interpolation support are valid. A projection-screen event is bracketed by consecutive saved states and does not alone establish terminal annihilation. For the operational screen used in the recorded trajectories, the minimum tangent norm must exceed 0.05 in both occupied-cell and reconstructed-surface diagnostics.

Velocity fits use only valid pre-event positions. The high-current comparison uses the common nominal interval 0.04--0.30 ns; additional fitting-window comparisons are given in Supplementary Section~\ref{sec:S5}, and the actual sample endpoints are included in the underlying data available as described in the main article's \hyperref[sec:data-availability]{Data availability statement}. The low-current late velocity uses the final 100 ps of the measured trajectory. These conventions separate finite-time motion from a fitted steady mobility. Train registration is each bit's displacement relative to the instantaneous mean displacement of all four textures.
\subsection*{Passive readout}
External stray fields are reconstructed from saved magnetization, cell volumes, material maps and the appropriate finite or periodic sources. They are not identified with the internal effective field. Independent cuboid, subdivision, image-count and footprint checks are detailed in Supplementary Section~\ref{sec:S9}.

For a weak axial probe current, the anisotropic magnetoresistance model is \(\rho_{\rm el}=\rho_\perp+\Delta\rho\,m_s^2\) \citep{McGuire1975AMR, CorteLeon2014AMR}. We report normalized responses with uniform cross-sectional weighting and a 40 nm axial averaging window. A separate orientation proxy averages \((m_{s,\rm bg}-m_s)/2\), where the background is evaluated for the same geometry. Periodic windows wrap across the simulation seam; finite-track windows use only supported detector positions. No resistance coefficient, electrical transfer function, detector noise or back-action is imposed. The resulting signals are passive contrast estimates.

%% file: sections/S1.tex
\section{Geometry, surface charge and projected winding}\label{sec:S1}
Take the straight-tube axis to be \(x=s\), with local basis \(\mathbf e_s=(1,0,0)\), \(\mathbf e_r=(0,\cos\phi,\sin\phi)\) and \(\mathbf e_\phi=(0,-\sin\phi,\cos\phi)\). The ordered surface coordinates \((\phi,s)\) have outward normal \(\mathbf e_r\). The shell occupies \(R_i<r<R_o\). For the unit field \(\mathbf m=a\mathbf e_r+b\mathbf e_\phi+c\mathbf e_s\), rotational symmetry means that the local components \(a,b,c\) are independent of \(\phi\); the Cartesian vector still rotates with the basis.

The surface topological charge is

\begin{equation}
\label{eq:charge}
Q=\frac{1}{4\pi}\int \mathbf m\cdot(\partial_\phi\mathbf m\times\partial_s\mathbf m)\,d\phi\,ds.
\end{equation}

The cylindrical metric factor in the area element cancels that in the physical circumferential derivative. For a rotationally symmetric unit field, \(\partial_\phi\mathbf m=\mathbf e_s\times\mathbf m\), and the triple product equals \(\partial_s c\). Hence

\begin{equation}
\label{eq:charge-endpoints}
Q[s_1,s_2]=\frac{c(s_2)-c(s_1)}{2}.
\end{equation}

An ideal complete 180° wall has \(Q=\pm1\) and a 360° pair has total \(Q=0\), consistently with the nanotube single-wall classification \citep{LeeKim2021Orbital}. End circles uniformly magnetized at an axial spin-sphere pole, \(+\mathbf e_s\) or \(-\mathbf e_s\), can each be identified with a point, permitting the degree construction on a capped cylinder. An axially periodic support instead has toroidal topology and still admits a degree. Free ends permit charge flux. During deformation, a dividing circle between constituents need not be pole-valued, so their partial charges need not remain separately quantized.

The tangent projection \(\psi=c+ib\) lies in a punctured complex plane whenever \(\psi\ne0\). Its phase winding is invariant under continuous deformation that avoids the origin and preserves the boundary closure. A useful tangent trial field is

\begin{equation}
\label{eq:trial}
\mathbf m=\sin\theta(s)\mathbf e_\phi+\cos\theta(s)\mathbf e_s,\qquad
\theta(s)=2\arctan e^{(s-X_1)/\Delta}+2\arctan e^{(s-X_2)/\Delta}.
\end{equation}

Here \(X_1,X_2\) are constituent positions and \(\Delta\) is the inverse-slope width of each isolated trial wall. The sum gives a same-handed full turn. Replacing the sum by a difference gives a retracing pair with zero winding. These one-dimensional spin trajectories can resemble a diameter through other localized textures; they do not determine the full two-dimensional surface charge by themselves.

\subsection{A smooth finite-energy escape}\label{a-smooth-finite-energy-escape}

For a tangent same-handed field, define a path parameter \(0\le\lambda\le1\) and

\begin{equation}
\label{eq:contraction-vars}
\xi=(1-\lambda)\cos\theta+\lambda,\quad
\zeta=(1-\lambda)\sin\theta,\quad
D_0=1+\xi^2+\zeta^2,
\end{equation}

\begin{equation}
\label{eq:contraction}
(a,b,c)=\frac{1}{D_0}\left(\xi^2+\zeta^2-1,\ 2\zeta,\ 2\xi\right).
\end{equation}

At \(\lambda=0\) this is the original tangent field, and at \(\lambda=1\) it is uniformly \(+\mathbf e_s\). The denominator is at least one and \(a^2+b^2+c^2=1\). At \(\theta=\pi\), \(\lambda=1/2\), the spin points along \(-\mathbf e_r\): the tangent projection vanishes but the full magnetization remains regular. The Cartesian field \((c,a\cos\phi-b\sin\phi,a\sin\phi+b\cos\phi)\) closes smoothly around the circumference and keeps the axial asymptotes fixed.

The same components extend through a hollow finite-thickness shell because \(R_i>0\). For a smooth localized trial profile, \(|\partial_s\mathbf m|^2\le4\theta'^2\) and \(|\partial_\phi\mathbf m|^2=a^2+b^2\), so the exchange integral is finite, including its \(\ln(R_o/R_i)\) radial factor. Local quadratic anisotropy has finite excess. Relative to the axial background, the localized magnetization change is square-integrable; the longitudinal-projection bound \(E_d(\delta\mathbf M)\le\mu_0\|\delta\mathbf M\|_2^2/2\) establishes finite magnetostatic energy \citep{DiFratta2020Variational}. This construction proves the availability of a nonsingular escape, not its minimum barrier. It parallels the phase-slip logic of other constrained magnetic order parameters \citep{Kim2016PhaseSlips}.

\subsection{Circumferential background}\label{circumferential-background}

For a background \(+\mathbf e_\phi\), the Cartesian magnetization varies around the end circles. Those boundary maps must be retained rather than compactified as constant axial domains. A relative degree can be defined by comparison with a fixed reference interpolation. The local transformation \((a,b,c)\mapsto(a,c,-b)\) maps the contraction above to one ending at \(+\mathbf e_\phi\). The background has nonzero exchange per unit length, so finite-energy statements refer to excess energy. Circumferential 180° axial-core rotations have zero total surface charge and different partial-charge sectors and magnetostatic interactions from the axial branch. The nonbinding examples studied for that family therefore do not contradict binding of the axial pair.

%% file: sections/S2.tex
\section{Energetic binding and static resolution}\label{sec:S2}
The binding calculation separates two roles. Rotation sense fixes whether approaching walls can remove their winding without a radial excursion. The energy determines whether their separation is finite. Opposite axial-wall poles attract at large distance, whereas overlapping same-handed profiles repel. The energy minimum and return after perturbation test these ingredients separately.
In the local cylindrical basis, exchange is

\begin{equation}
\label{eq:exchange}
|\nabla\mathbf m|^2=|\partial_r(a,b,c)|^2+|\partial_s(a,b,c)|^2+
\frac{(\partial_\phi a-b)^2+(\partial_\phi b+a)^2+(\partial_\phi c)^2}{r^2}.
\end{equation}

For a radially uniform tangent field, radial integration gives a curvature contribution \(AC_g\sin^2\theta\), where \(C_g=2\ln(R_o/R_i)/(R_o^2-R_i^2)\). With shell area \(S=\pi(R_o^2-R_i^2)\) and energy scale \(E_0=AS/\Delta_0=1.306902544\times10^{-18}\) J, the local model is

\begin{equation}
\label{eq:local-energy}
E_{\rm loc}/S=\int[A\theta'^2+K_{\rm eff}\sin^2\theta]ds,\qquad
K_{\rm eff}=K_\phi+AC_g.
\end{equation}

Its finite-energy stationary solution satisfies \(A\theta'^2=K_{\rm eff}\sin^2\theta\). Reaching \(\theta=\pi\) takes infinite axial distance, so this unbiased local model does not support a finite-separation monotone 2π wall. Same-handed overlap is repulsive \citep{Muratov2008Theory360, Braun1994Pairs}.

The axial constituents carry opposite effective magnetic charges of magnitude \(q_m\simeq2M_sS\). Their far-separated magnetostatic interaction is attractive, \(-\mu_0q_m^2/(4\pi d)\), with constituent separation \(d\). The finite-thickness trial calculation evaluates nonlocal magnetostatics beyond this asymptote and finds a minimum near \(d/\Delta_0=5.134\), with \(\Delta_0=20\) nm. Unconstrained relaxation yields \(d=101.750\) nm and inverse-slope widths around 20.6 nm. Retracing controls can unwind within the tangent manifold; same-handed interior annihilation requires leaving the nonzero-projection class.

The selected fine-grid endpoint has \(d=101.851\) nm, width larger by 0.165\% and excess energy lower by 1.407\%. Its raw maximum transverse effective field, \(\max_{i\in\Omega}|\mathbf m_i\times\mathbf B_{{\rm eff},i}|\) over occupied cells, is about 0.500 mT, compared with 0.0637 mT for the coarse reference. Signed perturbations and independent initial separations support the discrete local basin. These checks constrain static model sensitivity but do not supply a continuum error bar for every observable. The fourfold component of the residual decreased strongly with near-state minimization, so the remaining error is not identified as an irreducible staircase floor.

The reference's circumferential penalty and its full magnetostatic field are both essential parts of the stated energy model. An additional radial material anisotropy is a distinct intervention, not a second copy of shape anisotropy.

\subsection*{Restoration after small unconstrained perturbations}
The ten perturbations are applied to the same relaxed axial pair, with coefficients \(\epsilon=\pm0.01\). For each trial field \(\mathbf v\), we remove its spin-parallel component using the projection \(\mathcal P_{\mathbf m}\) defined in Section~\ref{sec:S4}, set vacuum cells to zero, and divide by its maximum occupied-cell norm. The perturbed state is
\begin{equation}
\mathbf m_\epsilon=\frac{\mathbf m+\epsilon\widehat{\mathbf v}}{|\mathbf m+\epsilon\widehat{\mathbf v}|},\qquad
\widehat{\mathbf v}=\frac{\mathcal P_{\mathbf m}\mathbf v}{\max_\Omega|\mathcal P_{\mathbf m}\mathbf v|}.
\end{equation}
The maximum vector change is 0.0099996 (approximately a 0.01 rad rotation), not a field amplitude. With \(f(s)=\exp[-(s/(8\Delta_0))^4]\), \(h(s)=\tanh(s/\Delta_0)\), and \(r\) the radius, the five applied fields before this normalization are
\begin{align}
\mathbf v_r&=fh\mathbf e_r,&
\mathbf v_\eta&=fh(\mathbf e_s\times\mathbf m),\\
\mathbf v_\phi&=f[-\Delta_0\partial_s\mathbf m\cos\phi+\tfrac12\mathbf e_r\sin(2\phi)],&
\mathbf v_t&=f[2(r-R_i)/(R_o-R_i)-1]\mathbf e_r.
\end{align}
The fields $\mathbf v_r$, $\mathbf v_\eta$, $\mathbf v_\phi$ and $\mathbf v_t$ define radial tilt, core-spin rotation, circumferential variation and through-thickness variation, respectively. For combined distortion, the trial field is a spatially smoothed Gaussian random vector field multiplied by \(f\), with seed 317 and Gaussian widths one transverse cell and two axial cells. The axial derivative in \(\mathbf v_\phi\) is the centered periodic difference. The core-spin rotation (helicity) perturbation sign is an applied vector-field convention, separate from the measured helicity convention in Section~\ref{sec:S4}. All components remain free during subsequent relaxation; no perturbation field is maintained.

The return distance in Fig.~\ref{fig:escape}a is the occupied-volume RMS spin-angle difference,
\begin{equation}
\delta\theta_{\rm rms}=\left[\frac{1}{N_\Omega}\sum_{i\in\Omega}\operatorname{atan2}^{2}\!\left(|\mathbf u_i\times\mathbf u_i^{\rm ref}|,\mathbf u_i\cdot\mathbf u_i^{\rm ref}\right)\right]^{1/2},
\end{equation}
reported in degrees. Here $\mathbf u_i$ and $\mathbf u_i^{\rm ref}$ are unit vectors after registering only one common periodic axial translation by minimizing the full spin-vector RMS difference. $N_\Omega$ counts all occupied cells. Separation, helicity, azimuthal and thickness distortions remain in the comparison. For small differences this angular RMS, in radians, agrees with the vector-norm RMS.

All ten initial and final fields use the same unperturbed reference, obtained by near-state minimization followed by a bounded 240 s wall-clock relaxation verification on the $4\times1\times1$ nm mesh. Its separation is 101.743902 nm. This common reference is used for both the eight shorter perturbation tests and the two extended tests of circumferential variation; the binding and mesh comparisons above use their separately relaxed endpoints. The initial angles consequently include the small change of the unperturbed state during subsequent relaxation.

To assess reference sensitivity, two independent solver continuations start from the identical saved unperturbed state. One uses \texttt{Relax} for 1080 s; the other uses \texttt{Minimize}, which returns after about 170 s, followed by 240 s of \texttt{Relax}. These durations are wall-clock times. Both retain the geometry, energy model, full demagnetization and zero drive. The settings are \texttt{MaxErr} $=10^{-7}$, \texttt{MinimizerStop} $=10^{-7}$, \texttt{MinimizerSamples} $=10$ and \texttt{RelaxTorqueThreshold} $=10^{-5}$ T. These are solver-termination settings, not angular error bars. The two final fields differ by 0.002348° RMS, principally through a separation difference of 0.005889 nm. Their final checkpoint-to-endpoint changes are only $1.35\times10^{-6}$° and $8.28\times10^{-7}$°, respectively, yet their protocol-dependent difference persists. Both relaxation calls reach their time limits and retain maximum transverse effective fields near 0.0633 mT. This protocol dependence does not establish an angular precision or confidence band.

The tested synthetic fractional-cell translation round trips leave at most $4.7\times10^{-7}$° RMS registration error. Tightening the registration optimizer changes the audited angular differences by less than $10^{-10}$°. A componentwise float32 estimate gives a single-field RMS encoding-error bound of $2.6\times10^{-6}$° for the audited reference, assuming nearest rounding. These checks separate registration and storage effects from relaxation and field-evaluation error; they do not bound continuum interpolation or mesh uncertainty.

All ten original endpoints retain two wall crossings, $W=1$, near-zero total $Q$ and a nonzero tangent projection. Their remaining angular differences exceed the unperturbed solver-method difference. A linearized separation mode accounts for more than 99\% of the squared registered spin-vector residual in eight cases; the two cases of circumferential variation retain additional shape deformation. This diagnostic does not remove separation from the plotted distance. Applying the same minimization--relaxation sequence to the positive circumferential-variation endpoint reduces its difference from the common reference to 0.01160°, while its last verification interval still changes by $1.74\times10^{-5}$°. This additional control is reported separately and does not replace one member of the original ten-test comparison. The results establish a restoring response with small resolved residual distortions, rather than exact return within numerical precision. Numerical relaxation progress is not physical elapsed time.

Figure~\ref{fig:perturbation-vectors} separates the applied change from the actual magnetization for the positive radial-tilt test. Here $A$ is the exact saved state to which the perturbation was applied, $C$ is the perturbed state and $C'$ its relaxed endpoint. The displayed difference fields are $B=C-A$ and $D=C'-A$, including the unit-normalization step in $B$. Their arrow sizes use one global amplification factor per field, whereas the accompanying maps share an absolute scale. No registration is applied to these visual differences. Relative to $A$, the occupied-volume RMS angle falls from 0.26131° to 0.01832°. The remaining difference is dominated by a small change in constituent separation; the RMS radial component of the difference falls from $4.56\times10^{-3}$ to $1.07\times10^{-5}$. This comparison includes any continued relaxation of $A$; it differs from the later common-reference, translation-registered comparison in Fig.~\ref{fig:escape}a and Table~\ref{tab:perturbations}.

Figures~\ref{fig:perturbation-radial}--\ref{fig:perturbation-mixed} show $B$ and $D$ for both signs of each perturbation type, using the same names as Fig.~\ref{fig:escape}a and Table~\ref{tab:perturbations}. The actual textures $A$, $C$ and $C'$ remain visually similar at unit scale, so their full sequence is shown only for the radial-tilt example. Each difference field uses its own explicitly labelled global amplification; the accompanying magnitude maps share the absolute $[0,0.01]$ scale. Through-thickness variation is sampled separately near the inner and outer shell surfaces at $r=78.8$ and $81.2$ nm, with one shared scale for both layers of a given field. The extended circumferential-variation endpoints are retained; they are not replaced by the later polishing control.

\begin{table}[htbp]\centering
\caption{Signed perturbation returns relative to the single unperturbed reference defined above. Both RMS angles include all occupied cells after translation-only registration. Energy and separation differences use this same reference; reported energy differences are comparable to the few-$10^{-6}E_0$ evaluation scatter seen in the solver controls and do not establish field identity. All ten retain $W=1$, near-zero total $Q$ and nonzero tangent projection.}\label{tab:perturbations}
\setlength{\tabcolsep}{3pt}
\begin{tabular}{lrrrrr}\toprule
Mode & $\epsilon$ & Initial RMS (°) & Final RMS (°) & $\Delta d$ (nm) & $\Delta E/E_0$ ($10^{-6}$)\\\midrule
Radial tilt & -0.01 & 0.26382 & 0.01833 & +0.04648 & +0.77 \\
Radial tilt & +0.01 & 0.26311 & 0.01528 & +0.03876 & +1.53 \\
Core-spin rotation & -0.01 & 0.15775 & 0.01506 & +0.03820 & +0.00 \\
Core-spin rotation & +0.01 & 0.15775 & 0.01683 & +0.04269 & +0.77 \\
Circumferential variation & -0.01 & 0.15946 & 0.01689 & -0.01156 & +0.77 \\
Circumferential variation & +0.01 & 0.15946 & 0.01703 & -0.01136 & +0.00 \\
Through-thickness variation & -0.01 & 0.16544 & 0.01577 & +0.04000 & +1.53 \\
Through-thickness variation & +0.01 & 0.16544 & 0.02189 & +0.05551 & +2.30 \\
Combined distortion & -0.01 & 0.09151 & 0.02376 & +0.06009 & +2.30 \\
Combined distortion & +0.01 & 0.09184 & 0.01405 & +0.03556 & +0.77 \\
\bottomrule\end{tabular}\end{table}

The nominal starting separations are 60.0, 102.686 and 160.0 nm. The 60 nm overlapping profile has a measured initial zero-crossing separation of 64.49 nm. The final short, reference and wide branches have separations 101.602, 101.750 and 101.891 nm. The two alternative endpoints differ from the reference by volume RMS distances 0.001023 and 0.000966, and by \(1.53\times10^{-6}E_0\) and \(3.06\times10^{-6}E_0\). Figure~\ref{fig:robustness}a compares the three starts; Figs.~\ref{fig:stability} and~\ref{fig:perturbation-maps} show the full same-handed/retracing component maps and the radial-tilt example. These bounded tests support a restoring local basin; they do not measure the largest survivable perturbation or a thermal retention time.

\subsection*{Circumferential background as a physical comparison}
For a circumferential background the easy direction is \(\pm\mathbf e_\phi\), which is uniform only in the local frame. A separately implemented negative circumferential coefficient provides this easy axis; its reference value is approximately \(-34.53\) kJ/m³. Its ideal tangent 180° wall reverses through an axial core and carries no corresponding \(\pm1\) surface-charge sector (Section~\ref{sec:S1}). The leading long-range interaction between the axial core moments of a same-handed pair is repulsive, proportional to \(+d^{-3}\), instead of the axial-background pole attraction. The tested circumferential pair expands, including a shifted-placement control from about 140.90 to 154.42 nm, without establishing a localized bound counterpart. Retracing configurations relax away. This is a result for the sampled energy model and interval, not a prohibition on circumferential-background binding with other interactions. The comparison separates the influence of background orientation and magnetostatics from projected winding alone.

\subsection*{From an internally bound pair to a transported train}
The attraction between the opposite leading poles of two constituent walls competes with their short-range, same-handed overlap repulsion, selecting the internal separation of one 360° wall. This does not select the spacing between different 360° walls. At distances $L\gg R,d$, identical isolated composites with parallel axial excess moment $p$ interact to leading order as dipoles, $U_{\rm inter}\simeq-\mu_0p^2/(2\pi L^3)$ (Supplementary Section~\ref{sec:S9}), which decays faster than the constituent pole interaction. At closer spacing, profile overlap and deformation also matter; their full inter-composite force law has not been measured here. The periodic train therefore tests finite-time preservation of internally bound carriers, rather than an equilibrium bit pitch.

%% file: sections/S3.tex
\section{Localized escape and the barrier estimate}\label{sec:S3}
The transition configuration on the lowest-cost route found lies at \(0.293483E_0=2.393950\) eV above the prepared pair, where \(E_0=1.306902544\times10^{-18}\) J. Its maximum absolute radial component, \(\max_{i\in\Omega}|m_{r,i}|\), is 0.9593, minimum tangent norm 0.28235 and angular participation measure approximately 0.189. Constituent separation is about 83.92 nm. The sampled thickness RMS is 0.000353 at this configuration; its maximum across the 42 inspected states is 0.000531. Total surface charge stays near zero. The largest adjacent-spin angle is 13.68° at this configuration and about 16.39° over the inspected sequence. These measures support a resolved smooth route on the available mesh.

The angular and thickness diagnostics use normalized bilinear spin samples at $N_\phi=192$ equally spaced angles and three radii $r_k=R_i+(R_o-R_i)\{0.2,0.5,0.8\}$, at all $N_x$ axial cell centers. Interpolation accounts for occupied-cell coverage. Define the angular radial-spin power and participation by
\begin{equation}
\label{eq:angular-participation}
 P_j=\frac{1}{3N_x}\sum_{k=1}^{3}\sum_x m_r(r_k,x,\phi_j)^2,
 \qquad \mathcal A=\frac{(\sum_jP_j)^2}{N_\phi\sum_jP_j^2}.
\end{equation}
This normalized inverse participation ratio is one for uniform angular power and decreases with localization; it is not a thresholded angular fraction. There is no wall-only mask or radial Jacobian weight. It is unavailable for $\sum_jP_j\leq10^{-20}$. The thickness measure is
\begin{equation}
\label{eq:sampled-thickness-rms}
 T_{\rm rms}=\left[\frac{1}{3N_xN_\phi}
 \sum_{x,j}\sum_{a\in\{r,\phi,s\}}
 \big(m_a(r_{0.8},x,\phi_j)-m_a(r_{0.2},x,\phi_j)\big)^2\right]^{1/2}.
\end{equation}
It is a componentwise RMS between the 80\% and 20\% shell-depth samples, over the full axial and angular support. It is neither a thickness derivative nor a vector-norm RMS, and alone does not establish continuum through-thickness convergence.

The energy changes are approximately +2.712 eV exchange, +2.486 eV demagnetization and −2.805 eV circumferential anisotropy. The residual transverse field \(\mathbf B_\perp=\mathbf B_{\rm eff}-(\mathbf m\cdot\mathbf B_{\rm eff})\mathbf m\) has RMS magnitude \(\langle|\mathbf B_\perp|^2\rangle_\Omega^{1/2}=0.336\) mT and maximum magnitude 2.266 mT, with equal-volume averaging over occupied cells. One unstable-direction curvature is approximately \(-93E_0\)/rad²; a half-step refinement changes the energy by about −0.000144 eV (see the \hyperref[sec:saddle-local-checks]{directional-curvature and step-size checks below}). Full saddle index, global pathway optimality and spatial convergence have not been established \citep{Bessarab2015GNEB, E2007String}.

Energy descent on opposite sides approaches the pair and the uniform state. A sampled low-energy connection between the returning configuration and the reference supports a common basin. The coherent comparison reaches about 7.27 eV but is not saddle-converged. Thus localization reduces the sampled cost; neither value is certified as the globally minimal activation barrier. Path ordering is not physical time.

The tangent projection remains nonzero at the transition configuration. Its winding changes during descent toward the uniform state. The reverse route from the uniform state includes the approximately 76.3 eV formation energy of the pair and reaches about 78.7 eV at this saddle. Neither reverse-route energy nor the forward 2.4 eV estimate is electrical writing energy. The reference barrier is not transferred to thicker tori or added-radial-anisotropy models.

\paragraph{Directional-curvature and step-size checks.}\label{sec:saddle-local-checks}
The local path tangent is the difference of the spin-sphere logarithms from the transition state to its two neighboring path configurations, normalized so that \(\langle|\mathbf t_i|^2\rangle_\Omega=1\) over occupied cells. A displacement \(q\) rotates each spin along its great circle through angle \(q|\mathbf t_i|\); thus \(|q|\) is the RMS spin-rotation angle. The directional curvature is evaluated as \([E(q)+E(-q)-2E(0)]/q^2\), using \(|q|=0.008\), 0.016 and 0.0213267 rad; the quoted value uses 0.008 rad. For the step-size check, the climbing-image optimization is restarted with both the initial and maximum numerical integration steps of the fast inertial relaxation engine (FIRE)~\citep{Bitzek2006FIRE} halved. These are optimizer parameters; the magnetic mesh and the two neighboring path configurations remain fixed.

%% file: sections/S4.tex
\section{Collective-coordinate transport}\label{sec:S4}
For coordinates \(q_i\), the gyrotropic tensor \(G_{ij}\) and the dissipative tensor \(D_{ij}\) can be written as

\begin{equation}
\label{eq:tensors}
G_{ij}=\frac{M_s}{\gamma}\int\mathbf m\cdot(\partial_{q_i}\mathbf m\times\partial_{q_j}\mathbf m)dV,
\qquad
D_{ij}=\frac{M_s}{\gamma}\int\partial_{q_i}\mathbf m\cdot\partial_{q_j}\mathbf m\,dV.
\end{equation}

The helicity generator is \(\partial_\eta\mathbf m=-\mathbf e_s\times\mathbf m\), localized to each constituent. Reversing this angle convention reverses the associated tensor entries and angle rates together. Projecting the stated spin-transfer equation gives

\begin{equation}
\label{eq:projected-dynamics}
(\alpha D-G)\dot{\mathbf q}=-\nabla_{\mathbf q}U+(\beta D-G)\mathbf e_Cu,
\end{equation}

where \(U\) is the collective-coordinate energy and \(\mathbf e_C\) selects translation \citep{Tretiakov2008Collective, Thiaville2005STT}. For \(\mathbf q=(C,d,\bar\eta,\delta\eta)\), \(C=(X_1+X_2)/2\), \(d=X_2-X_1\), \(\bar\eta=(\eta_1+\eta_2)/2\), \(\delta\eta=\eta_2-\eta_1\) and \(g=2M_sS/\gamma\), the separated-wall approximation gives

\begin{equation}
\label{eq:gyro-components}
G_{C,\delta\eta}=G_{d,\bar\eta}=-g,\qquad G_{C,\bar\eta}=0,
\end{equation}

with dissipative diagonal \((2g/\Delta,g/(2\Delta),2g\Delta,g\Delta/2)\). Therefore

\begin{equation}
\label{eq:translation}
\frac{2\alpha g}{\Delta}\dot C+g\dot{\delta\eta}
=F_C+\frac{2\beta g}{\Delta}u,
\end{equation}

where \(F_C=-\partial_CU\). Dividing by \(2\alpha g/\Delta\) gives explicitly
\begin{equation}
\dot C=\frac{\beta}{\alpha}u-\frac{\Delta}{2\alpha}\dot{\delta\eta}+\frac{\Delta}{2\alpha g}F_C.
\label{eq:translation-bridge}
\end{equation}
For \(F_C=0\) this reduces to the main-text relation with \(c_\eta=\Delta/(2\alpha)\) in the separated limit.

Tensor integration on the actual reference field instead gives normalized \(G_{C,\delta\eta}=-0.957815\) and \(D_{CC}=2.094391\), in coordinates scaled by \(\Delta_0=20\) nm and tensor unit \(g\Delta_0\). Retaining the small off-diagonal dissipative terms gives \(c_\eta=91.43286\) nm/rad. In the coordinate order \((C/\Delta_0,d/\Delta_0,\bar\eta,\delta\eta)\), the complete dimensionless matrices (entries below \(2\times10^{-8}\) rounded to zero) are
\begin{equation}
\frac{G}{g\Delta_0}=\begin{pmatrix}0&0&0&-0.957815\\0&0&-0.957815&0\\0&0.957815&0&0\\0.957815&0&0&0\end{pmatrix},
\end{equation}
\begin{equation}
\frac{D}{g\Delta_0}=\begin{pmatrix}2.094391&0&0&-0.003346\\0&0.469205&-0.003346&0\\0&-0.003346&1.955600&0\\-0.003346&0&0&0.458221\end{pmatrix}.
\label{eq:integrated-tensors}
\end{equation}
The voxel-integrated unit is \(g\Delta_0=1.471178741\times10^{-28}\) J s. These are approximate coordinate derivatives, not fitted eigenmodes. Replacing the axial spectral derivative with the centered native stencil changes the gyro entry to −0.95139 and \(D_{CC}\) by −1.19\%, preserving the coupling pattern.

\subsection*{Helicity estimator and reference tangent basis}
For the periodic straight tube, $s=x$, $\phi=\operatorname{atan2}(z,y)$ and $(m_r,m_\phi,m_s)$ are local cylindrical components. The measured pair center is the phase of the axial circular moment,
\begin{equation}
\label{eq:measured-center}
 C_{\rm circ}=\frac{L}{2\pi}\arg\sum_x w_x e^{2\pi i x/L},\qquad
 w_x=\frac{1}{N_\perp}\sum_{\text{occupied cross section}}(1-m_s),
\end{equation}
with $L=1{,}024$ nm. The center is unavailable if the moment magnitude is at most $10^{-8}$. With $\xi=\operatorname{wrap}_{(-L/2,L/2)}(x-C_{\rm circ})$ and $\Delta_0=20$ nm, define
\begin{equation}
\label{eq:helicity-partition}
 p_C(x)=\tanh\!\left[\frac{L}{2\pi\Delta_0}\sin\!\left(\frac{2\pi\xi}{L}\right)\right].
\end{equation}
The helicity estimator sums over all occupied magnetic cells, with equal cell volumes and no radial selection or Gaussian window:
\begin{equation}
\label{eq:measured-helicity}
\begin{aligned}
 Z_1&=\sum_v(m_\phi+i m_r)(1-m_s^2)(1-p_C), & \eta_1&=\arg Z_1,\\
 Z_2&=\sum_v(m_\phi+i m_r)(1-m_s^2)(1+p_C), & \eta_2&=\arg(-Z_2).
\end{aligned}
\end{equation}
The $+m_\phi$ and $-m_\phi$ tangent cores both have zero reference angle. Positive $\eta_1$ rotates the first core toward $+m_r$; positive $\eta_2$ rotates the second toward $-m_r$. The initial measured angle is not subtracted at each frame. The saved relative phase is $\arg\exp[i(\eta_2-\eta_1)]$. For the displacement balance, this sequence is temporally unwrapped with a $2\pi$ period and a $\pi$ jump threshold, and its change from the first sample is used. Continuation slope fits likewise unwrap the relative and arithmetic common-angle sequences.

The tensor calculation uses a different center: $C_{\rm ref}$ is the mean of the two zero crossings of the azimuthally averaged mid-radius $m_s$ in the static reference. Its fixed partition $p_{\rm ref}$ has the form of Eq.~\eqref{eq:helicity-partition} with this center. For scaled coordinates $\widetilde{\mathbf q}=(C/\Delta_0,d/\Delta_0,\bar\eta,\delta\eta)$, the integrated tangent vectors are
\begin{equation}
\label{eq:implemented-tangent-basis}
\begin{aligned}
 \mathbf v_C&=-\Delta_0\mathcal P_{\mathbf m}\partial_x\mathbf m,&
 \mathbf v_d&=-\frac{\Delta_0}{2}p_{\rm ref}\mathcal P_{\mathbf m}\partial_x\mathbf m,\\
 \mathbf v_{\bar\eta}&=-\mathbf e_s\times\mathbf m,&
 \mathbf v_{\delta\eta}&=-\frac{p_{\rm ref}}{2}\mathbf e_s\times\mathbf m,
\end{aligned}
\end{equation}
where $\mathcal P_{\mathbf m}\mathbf a=\mathbf a-(\mathbf m\cdot\mathbf a)\mathbf m$. The primary derivative is periodic spectral; the centered difference is a sensitivity check. Equal-volume occupied-cell sums of these vectors evaluate Eq.~\eqref{eq:tensors}. This approximate four-coordinate basis is distinct from the transverse-core weights in Eq.~\eqref{eq:measured-helicity}; it is not an independently fitted angle trajectory.

Over the first 0.5 ns, measured excess displacements beyond the constant-angle contribution are −9.294 nm for torque alone and −9.705 nm with the self-field. The tensor relation combined with measured relative helicity supplies −8.957 and −9.218 nm: 96.38\% and 94.98\%. The angle trajectory is measured, not independently predicted by this displacement balance. Later physical continuation gives final-window velocities −35.99967 and −35.99896 m/s, compared with −35.9034 m/s from \(\beta u/\alpha\). Residual angle rates and the approximately 0.6\% advection-discretization error account for an uncertainty on the scale of tenths of a metre per second. Fits with roughly 0.35 ns decay times describe the uniform-drive transient.

A uniform axial field gives the leading pair excess \(2M_sSB_sd\), so \(F_C=0\) and \(F_d=-2M_sSB_s\). It primarily loads the separation rather than providing uniform translation pressure in the homogeneous straight model. This is selective translational immunity, not immunity to field-induced deformation or erasure.

The settled \(\beta u/\alpha\) spin-transfer velocity also appears in planar 360°-wall collective-coordinate theory \citep{LiLu2021Survey}. Here the independently measured internal-angle contribution and the full-vector radial escape provide the tubular comparison.

\subsection*{Response immediately after current switch-on}
For an initially stationary texture and torque-only axial drive, solving Eq.~\eqref{eq:main-LLG} at $t=0^+$ gives
\begin{equation}
(1+\alpha^2)\partial_t\mathbf m=-(1+\alpha\beta)u\,\partial_s\mathbf m
+(\beta-\alpha)u\,\mathbf m\times\partial_s\mathbf m.
\label{eq:initial-current-response}
\end{equation}
Thus the initial advective component is $(1+\alpha\beta)u/(1+\alpha^2)\simeq u$, accompanied by internal rotation; once the angles settle, translation approaches $(\beta/\alpha)u$. With the chosen $\beta/\alpha=1/2$, this accounts for the approximately twofold change in speed without an inertial acceleration law. In the saved torque-only trace the first 0--5 ps displacement gives $-70.726$ m/s. The Oersted trace starts at $-73.061$ m/s and has a resolved small additional maximum speed of 75.295 m/s (signed velocity $-75.295$ m/s) near 25 ps. The velocity estimated from the first 0--5 ps displacement is plotted at 2.5 ps in Fig.~\ref{fig:transport}c. 

Changing anisotropy can alter the angle excursion, its relaxation time and the accessible instability boundary. It does not by itself change the fixed-angle homogeneous-track relation $v=\beta u/\alpha$. Changing $\beta/\alpha$ or maintaining internal dynamics is a different intervention. Moderate internal-angle settling and eventual radial escape use related coordinates but are distinct outcomes; the low-current state settles intact.

%% file: sections/S5.tex
\section{Velocity range and comparison with magnonic transport}\label{sec:S5}
The common-window speeds below use the nominal 0.04--0.30 ns interval, restricted to valid pre-event positions. Conventional positive axial current gives negative velocity; the table gives magnitudes.

{\def\LTcaptype{table} 
\begin{longtable}[]{@{}
  >{\raggedright\arraybackslash}p{(\linewidth - 6\tabcolsep) * \real{0.2308}}
  >{\raggedright\arraybackslash}p{(\linewidth - 6\tabcolsep) * \real{0.2308}}
  >{\raggedleft\arraybackslash}p{(\linewidth - 6\tabcolsep) * \real{0.3077}}
  >{\raggedright\arraybackslash}p{(\linewidth - 6\tabcolsep) * \real{0.2308}}@{}}
\toprule\noalign{}
\begin{minipage}[b]{\linewidth}\raggedright
Current density (\(10^{12}\) A/m²)
\end{minipage} & \begin{minipage}[b]{\linewidth}\raggedright
Self-field
\end{minipage} & \begin{minipage}[b]{\linewidth}\raggedleft
Common-window speed (m/s)
\end{minipage} & \begin{minipage}[b]{\linewidth}\raggedright
Later observation
\end{minipage} \\
\midrule\noalign{}
\endhead
\bottomrule\noalign{}
\endlastfoot
1 & Off / on & 57.8 / 58.0 & Late speed approximately 36.0 m/s \\
2 & On & 116.8 & Intact to 1.2 ns \\
4 & On & 233.2 & Intact to 0.9 ns \\
8 & On & 472.8 & Intact to 0.7 ns \\
8 & Off & 464.9 & Matched 0.7 ns torque-only control \\
12 & On & 725.1 & Projection loss near 0.509 ns; terminal fate unresolved \\
16 & On & 1,006.8 & Projection loss near 0.376 ns; no pair by 0.411 ns \\
\end{longtable}
}

\subsection*{Equal-coefficient sign and advection control}
With \(\alpha=\beta=0.1\), \(P=0.4\) and torque-only drive, independently initialized positive and negative currents \(J_s=\pm10^{12}\) A/m² translate the reference by \(\mp14.16769\) nm over 0.20 ns. Fits to the final half of those intervals give \(-70.84059\) and \(+70.83992\) m/s, compared with the signed advection prediction \(\mp71.27366\) m/s. Their speed defect is about 0.61\%; the translated-reference vector RMS deformation is 0.000169 and final separation is 101.7108 nm. This validates the implemented sign and approximate advection of the discrete field. This control does not quantify dynamic mesh convergence. The \(1+\beta^2\) factor in Eq.~\eqref{eq:advection} means this prediction differs slightly from the \(\beta=0.05\) coefficient used for the internal-motion series.

Earlier and later fitting windows quantify transient sensitivity; for example, the highest-current speed is about 1,038 m/s over 0.04--0.14 ns. These fit variations are not statistical confidence intervals. The finite-ended 180° and 360° controls travel −53.861 and −53.473 nm over a matched 1.155189 ns, with final-window velocities −39.263 and −37.745 m/s. End forces and different preparation histories preclude a precision intrinsic-mobility ranking.

For a uniformly translating perturbation in an otherwise stationary background, the basic radiation condition is \(\omega_n(k)=vk\), where \(\omega_n(k)\) is the angular-frequency dispersion of an available spin-wave branch and \(k\) its axial wave number. A minimum positive phase velocity over coupled modes provides a Cherenkov-like onset scale. In the field-driven nanotube calculations of Yan et al., reaching that scale produces spin-wave emission and a strong change in mobility; propagation beyond onset is possible \citep{Yan2011FastNanotubes, Hertel2016Ultrafast}. The relevant scale is material-, geometry-, mode- and drive-dependent, not a universal number. A current-driven background may additionally shift its dispersion.

The present high-speed measurements establish motion in the kilometer-per-second range, but no radiation spectrum or dispersion was computed for this 360° pair. They therefore do not locate its own magnonic onset. Other high-speed approaches, including synthetic antiferromagnets and ferrimagnetic spin-transfer systems, provide important comparators \citep{Yang2015SAFFast, Wu2026SpinelDW}. The distinction under investigation here is the combination of a localized composite, hollow geometry and resolved internal-mode dynamics, rather than an absolute speed record.

%% file: sections/S6.tex
\section{Differential drive and curved tracks}\label{sec:S6}
A curved track asks whether coupled segments of one wall can remain coherent despite unequal path lengths and unequal local forcing. The complete torus removes physical end faces; changing its imposed current profile isolates part of the forcing mismatch without changing its magnetic curvature. Figure~\ref{fig:current-fields} shows the actual exported currents and the coordinate conventions.
For a torus of major radius \(R\), let \(\vartheta\) run along its circular centerline. At tube radius \(r\) and cross-sectional angle \(\phi\), \(\rho=R+r\sin\phi\) and \(s=R\vartheta\). The local physical distance along the tube is \(d\ell=(\rho/R)ds\). In the recorded convention, \(\phi=0\) is top, \(90°\) outside, \(180°\) bottom and \(270°\) inside. Thus a contour interval in \(s\) is not a local physical wall width. For the prepared \(R=704\) nm torus, inner/outer separations of 78.16/69.75 nm in \(s\) correspond to 69.28/77.68 nm physically.

The physical ideal-conductor current is \(J=C_J/\rho\), cross-section-normalized. If uncoupled local translation were proportional to current, its angular rate would scale as \(J/\rho\propto\rho^{-2}\). Constant \(|J|\) leaves a \(\rho^{-1}\) angular-rate variation; prescribed \(J\propto\rho\) compensates that scalar mismatch. Neither counterfactual removes the rotating spin frame or the magnetic curvature terms. Exported current vectors verify the implemented direction and amplitude: for the $10^{12}$ A/m$^2$ cross-sectional mean, the occupied-cell magnitude range is $0.88415$--$1.13361\times10^{12}$ A/m$^2$. The torus-volume mean is $0.99345\times10^{12}$ A/m$^2$, because volume weighting differs from the specified cross-sectional normalization.

\subsection*{Prescribed straight-tube drive and shape observable}
For the straight controls, $\mathbf e_s=\mathbf e_x$ and $\phi=\operatorname{atan2}(z,y)$, so $\phi=0$ points along $+y$ and $\pi/2$ along $+z$. This Cartesian convention differs from the torus labels above. The imposed current is
\begin{equation}
\label{eq:differential-current}
 \mathbf J(x,r,\phi)=\overline J\frac{f(\phi)}{\langle f\rangle_{\rm occ}}\mathbf e_x,
 \qquad f(\phi)=1-\epsilon\sin\phi,
 \qquad\langle f\rangle_{\rm occ}=\frac{1}{N_{\rm occ}}\sum_{v\in\Omega}f(\phi_v).
\end{equation}
Here $\overline J=-2\times10^{12}$ A/m$^2$ and $\epsilon=0.10$ or $0.25$. The mean is over equally weighted occupied magnetic cells; symmetric annular sampling gives $\langle f\rangle_{\rm occ}=1$. The ideal magnitude extrema are $(1\pm\epsilon)|\overline J|$, with lower magnitude at $+z$ and higher magnitude at $-z$. There is no explicit axial or through-thickness dependence. The separate one-sided control uses $f=1-\tanh[\sin\phi/0.20]$ with the same normalization. None of these straight profiles includes an Oersted field.

Wall contours use normalized trilinear spin samples at 96 equally spaced angles. The center harmonic uses the middle radius only. At each angle, exactly two periodic $m_s=0$ crossings are required: $l$ for $+\!\to\!-$ and $r$ for $-\!\to\!+$. Linear interpolation gives $d=(r-l)\bmod L$ and $C(\phi)=(l+d/2)\bmod L$. Centers are unwrapped about their angular circular mean into one interval of length $L$. With at least 75\% of angles valid, an equally weighted least-squares fit includes the intercept and the first four harmonics together:
\begin{equation}
\label{eq:wall-line-harmonics}
 C(\phi,t)=C_0(t)+\sum_{n=1}^{4}[a_n(t)\cos(n\phi)+b_n(t)\sin(n\phi)]+\text{residual}.
\end{equation}
The reported first-harmonic distortion subtracts coefficients of the first current-on sample from the same trajectory:
\begin{equation}
\label{eq:wall-line-amplitude}
 A_1(t)=\sqrt{[a_1(t)-a_1(0)]^2+[b_1(t)-b_1(0)]^2}.
\end{equation}
It has length units and is plotted in nanometres, without division by radius, mean travel or track length. It is the sinusoidal amplitude; a pure first-harmonic distortion has peak-to-peak variation $2A_1$. This is a within-trajectory shape reference, not subtraction of unmatched driven and zero-current histories.

Straight controls impose a first circumferential harmonic in axial current. For fractional modulation 0.10/0.25, the early center-harmonic rates are 11.382/28.458 m/s, with ratio 2.5003. The nominal fit interval is 0.002--0.177 ns. Changing its upper limit modifies the normalized slope, and later dynamics is nonmonotone. These results identify initial forcing linearity more securely than a unique restoring time or threshold.

At \(R=704\) nm and mean current magnitude \(2\times10^{12}\) A/m², the physical profile first fails the projection screen at 0.711 ns, between the last intact sample at 0.691 ns and that observation. Constant-magnitude drive gives an event near 1.010 ns; the angular-compensating control retains the pair for 201.1 nm. With physical current of \(10^{12}\) A/m², the 704 nm-radius trajectory is continued from its exact saved 3.53766 ns endpoint, with no intervening relaxation or minimization. The combined trajectory has 425 saved magnetic states and 547 scalar records and reaches 9.65866 ns. The last state passing the recorded integrity/projection screen occurs at 9.43766 ns, after 490.065 nm of centerline travel (39.884°, or 0.11079 of a circuit). The first failed screen at 9.45766 ns brackets the onset of an almost-radial excursion: the minimum tangent norm changes from 0.3415 to 0.03770, while winding is still one at that first failed sample. At the final frame, the minimum norm is 0.01625 and the sampled winding is no longer uniform around the circumference. Charge values obtained after interpolation support fails are not interpreted as valid topological measurements. Figure~\ref{fig:curvature} and Supplementary Movie~\ref{movie:torus} show both the intact interval and the late deformation.

The centerline-shear diagnostic remains modest through most of the continuation: 4.61 nm at 3.538 ns and 4.01 nm at 8.538 ns, increasing to 12.64 nm at 9.438 ns and 15.07 nm at the first failed screen. At the same late times the radial maximum rises sharply and the inner-side constituent spacing contracts. This supports a late developing deformation rather than an isolated anomalous final frame, without separating physical instability from numerical sensitivity.

This endpoint does not establish terminal annihilation, a numerical artifact or proximity to a quantitatively determined stability threshold. These calculations demonstrate approximately 490 nm of intact curved travel under the stated drive. No full circuit was simulated. The separate 1,056 nm-radius run retains the wall for 200.08 nm in 3.419 ns, with minimum tangent norm 0.949 and center shear 3.52 nm. At 200.95 nm of travel, the 704 nm-radius trajectory has norm 0.930 and shear 4.61 nm, providing a comparison with the 1,056 nm-radius trajectory over approximately 200 nm. A survivor at one current and an event at a different current do not isolate a radius boundary, and no monotonic radius threshold is inferred.

A radial penalty of \(0.25K_d\) preserves the straight 25\%-modulated state over 201.4 nm with minimum projection 0.943. Exchange increased by 20\% delays loss to 0.952 ns. No toroidal radial-anisotropy intervention or modified-model writing/barrier calculation was performed. The post-event torus/straight tails retain current; they do not establish current-off recovery or terminal annihilation in every event.

The finite 90°/180° modules had confounded incident-state preparation and did not demonstrate intact bend passage. A cropped-arc approximation failed its demagnetizing-field calibration and supplies no predictive large-radius result. These limits are distinct from the successful complete-torus finite-distance trajectories.

Unequal path lengths, current crowding and magnetic compensation have precedents in curved planar and ferrimagnetic tracks \citep{Garg2017Curvature, OsunaRuiz2021Curved}. Nunes et al.~study field-driven single walls in thinner bent tubes, with a different wall-width convention; Skoric et al.~analyze curvature-dependent wall-type stability \citep{Nunes2026BentTubes, Skoric2021CurvedNanotubes}. The different geometries, drives and observation intervals of these studies do not support a quantitative failure-threshold ranking against the present model.

\begin{table}[htbp]\centering\small
\caption{Finite observations in complete tori with the 14.1 nm shell. Current is the mean magnitude in $10^{12}$ A/m$^2$. ``Event'' means first failed projection screen; it is not automatically terminal annihilation. The two events share radius and mean current but change current profile. The low-current radius comparison is valid over the common first 200 nm; the extended 704 nm run adds a longer observation, not a radius threshold.}\label{tab:curved-outcomes}
\begin{tabular}{rrllp{43mm}}\toprule
$R$ (nm) & $|\overline J|$ & Profile & Time (ns) & Observed outcome\\\midrule
704 & 1 & Physical $1/\rho$ & 9.43766 & Intact through 490.065 nm (39.884°); first screen failure 9.45766 ns\\
1056 & 1 & Physical $1/\rho$ & 3.419 & Intact through 200.08 nm (10.86°)\\
704 & 2 & Physical $1/\rho$ & 0.691--0.711 & Last intact / first screen failure\\
704 & 2 & Constant magnitude & 1.010 & Projection event observed\\
704 & 2 & Compensating $\rho$ & 1.837 & Intact through 201.1 nm\\
\bottomrule\end{tabular}\end{table}

%% file: sections/S7.tex
\section{Encoded-pattern transport and recovery}\label{sec:S7}

The anisotropy-patch controls deliberately prescribe a spatial change in the material coefficient $K_\phi$, present from preparation onward, to test tolerance to a local inhomogeneity. They are not an effective anisotropy extracted from the texture's dipolar field: the nonlocal demagnetizing field remains a separate term. The signed Gaussian patches change the pre-existing circumferential hard axis by $\pm25\%$; neither patch changes it into an easy axis.
The binary representation \(101_{10}=1100101_2\) has seven data slots, four of which contain a 360° DW. An eighth empty slot separates periodic copies. With pitch 384 nm, the four initial occupied centers have relative positions \(0,p,4p,6p\), exposing nearest-neighbor spacings \(p,3p,2p\). Bit identities are tracked independently through the periodic seam; total winding four alone does not establish preservation of the word.

At \(-8\times10^{12}\) A/m², the four textures travel approximately 389--393 nm in 0.902 ns, with deviations from common displacement no larger than 2.44 nm. The tangent projection falls to about 0.356 after a further 20 ps off. Later high-drive branches can lose identity under continued driving or after current removal. Their branch checkpoints differ by 20 ps, so their difference is not an exactly matched causal subtraction.

The train protocols use spin torque alone, with no imposed Oersted field and full magnetic demagnetization retained. The lower-current pulse starts from the original prepared train and applies \(-4\times10^{12}\) A/m² for 2 ns, then evolves for 0.8 ns at zero current. Pulse-end displacements are 358.96, 348.14, 355.04 and 354.69 nm. Final displacements are 306.12, 287.65, 298.11 and 297.66 nm. The corresponding maximum deviations from mean displacement are 6.07 and 9.73 nm. At the end of the 0.8 ns current-off interval, constituent separations are 106.74, 106.94, 104.08 and 104.35 nm, with minimum tangent projection 0.97570. All four textures retain their order and identity. Internal recovery accompanies additional spacing drift and recoil; net travel in this lower-current example is less than one slot pitch.

Selected ±25\% Gaussian circumferential-anisotropy patches, with width parameter 22 nm, are crossed at \(4\times10^{12}\) A/m² without identity loss. Four separate exchange/anisotropy ±20\% tests retain a pair over 0.4 ns at \(10^{12}\) A/m². They demonstrate responses to these deterministic perturbations, not arbitrary-disorder immunity or statistical reliability.

\subsection*{Exact-checkpoint reverse continuation}
The 2.8 ns recovered checkpoint is continued for 1 ns at \(+4\times10^{12}\) A/m², with the same physical damping, precession and spin-transfer coefficients and no numerical relaxation. All four textures remain intact at 3.8 ns, with total projected winding four and minimum tangent norm 0.88944. Their reverse-phase displacements are \(-205.25,-215.69,-210.75,-210.89\) nm relative to their own 2.8 ns positions. These are finite-window reverse motions, not a fitted steady mobility. Supplementary Movie~\ref{movie:train} shows the complete forward/off/reverse sequence with one continuous physical-time origin.

\subsection*{Finite-time response to material nonuniformity}
After numerical preparation and a 20 ps zero-current physical segment, the patch and uniform material-sensitivity controls are driven with the phase clock reset to zero. They include the ideal annular Oersted field of their specified axial current. The local patch has \(K_\phi(s)=K_{\phi,0}[1+a\exp(-(s-512\,\mathrm{nm})^2/(2\sigma^2))]\), where \(a=\pm0.25\) and \(\sigma=22\) nm. The pair is prepared near \(s=640\) nm and driven toward decreasing \(s\) by \(J_s=4\times10^{12}\) A/m². The increased-anisotropy patch is traversed over 240.36 nm in 1.354 ns, ending at separation 91.48 nm and minimum tangent norm 0.936. The decreased patch gives 240.77 nm in 1.471 ns, separation 120.98 nm and norm 0.853. Figure~\ref{fig:robustness} distinguishes these local traversal tests from the separately prepared, spatially uniform \(A\) and \(K_\phi\) changes of \(\pm20\%\), each observed for 0.4 ns at \(10^{12}\) A/m². These observations do not establish a disorder-averaged reliability or a depinning threshold.

\begin{table}[htbp]\centering\small
\caption{Uniform material-parameter controls at $J_s=10^{12}$ A/m$^2$, each observed for 0.4 ns after its own preparation. All retain one pair, $W=1$ and valid nonzero projection; no perturbation ensemble or equilibrium phase boundary is inferred.}\label{tab:material-controls}
\begin{tabular}{lrrr}\toprule
Change & Travel magnitude (nm) & Final separation (nm) & Minimum tangent norm\\\midrule
$A-20\%$ & 22.95 & 88.31 & 0.9915 \\ 
$A+20\%$ & 22.96 & 114.19 & 0.9922 \\ 
$K_\phi-20\%$ & 22.08 & 110.53 & 0.9945 \\ 
$K_\phi+20\%$ & 23.98 & 94.54 & 0.9884 \\ 
\bottomrule\end{tabular}\end{table}

%% file: sections/S8.tex
\section{Localized field pulses}\label{sec:S8}
An ideal coaxial loop of radius 120 nm supplies its full axial and radial Biot--Savart field. Pulses have a 0.6 ns \(\sin^2\) envelope, with peak axial magnitude 0.25 or 0.40 T at the shell midpoint, followed by 0.4 ns with the field removed. The corresponding ideal loop currents are about 30.24 and 48.38 mA; the magnetic-shell transport current is zero. The writing tests start from the uniformly magnetized finite tube, and the erasure tests start from the prepared finite pair. No spin direction is directly overwritten during the pulse.

Both write cases return close to the uniform state, with no persistent target pair. The 0.25 T deletion attempt retains two crossings, separation 223.69 nm and winding near one. The 0.40 T attempt retains two crossings and separation 119.46 nm, with minimum projection 0.044 and mixed local turns; it does not produce a persistent uniform deletion target. These outcomes constrain the tested pulse family rather than all routes to writing or erasure. Relevant planar-wall manipulation and skyrmionium lifecycle demonstrations exist \citep{Zhang2016_360DW, Gartside2016Injection, Goebel2019SkyrmioniumRacetrack}. Experimental microcoils supply context for fields of several hundred millitesla, but do not validate the source geometry, waveform or efficiency assumed here \citep{Prasad2025Microcoils}.

A localized radial transition supplies a specific target for control. Spatially shaped fields can address the nucleus, differential torque can excite an internal distortion, and coherent excitation could build the relevant mode amplitude. Such selective coupling has experimental precedent in vortex-core switching \citep{VanWaeyenberge2006VortexReversal,Kammerer2011SpinWaveReversal}, but its efficiency and source feasibility for this tubular wall remain unmeasured. The present simple loop trials do not determine the performance of those protocols. The 0.25 T result demonstrates survival of a strongly distorted pair under this specified pulse; the near-radial mixed-winding 0.40 T outcome is neither an undamaged bit nor a demonstrated uniform erasure.

%% file: sections/S9.tex
\section{Stray-field and magnetoresistive contrast}\label{sec:S9}
For an observation point outside the magnetic region, the dipole-volume expression is

\begin{equation}
\label{eq:dipole}
\mathbf B(\mathbf r)=\frac{\mu_0}{4\pi}\int_\Omega
\frac{3\widehat{\mathbf R}[\mathbf M(\mathbf r')\cdot\widehat{\mathbf R}]-\mathbf M(\mathbf r')}{|\mathbf R|^3}dV',
\qquad\mathbf R=\mathbf r-\mathbf r'.
\end{equation}

The actual material mask, \(M_s\), vectors and cell volumes are used. Cuboid surface-charge comparison, source subdivision, periodic-image and footprint tests check this fixed-state reconstruction. The peak-normalized midpoint/subdivision difference is 0.0382\% at 40 nm gap and 0.0515\% at 30 nm; increasing the image count from ±4 to ±8 changes the sampled peak-normalized field by 0.00857\%. These are postprocessing checks, not total micromagnetic error estimates.

The prepared periodic pair gives approximately 10.22 mT peak vector contrast at 40 nm gap. Figure~\ref{fig:point-readout} shows the saved field scan and its constituent positions; the 5 nm-spaced scan has its largest vector contrast at $x=475$ nm. The field is reconstructed numerically from the simulated magnetization relative to its matched no-bit background. A 40 × 40 nm planar footprint parallel to the tube axis, sensitive along that axis, gives peak mean axial contrasts of 12.03, 5.83 and 1.85 mT at gaps of 30, 60 and 120 nm. Detector response, bandwidth, noise and magnetic back-action remain unspecified.

With uniform weak axial probe weighting, define

\begin{equation}
\label{eq:AMR}
L_{\rm AMR}=\int\left[\langle m_s^2\rangle_S-\langle m_{s,\rm bg}^2\rangle_S\right]ds,
\qquad\Delta R_{\rm AMR}=\frac{\Delta\rho}{S}L_{\rm AMR}.
\end{equation}

A tanh-profile 180° wall gives \(-2\Delta\); separated constituents give approximately \(-4\Delta\). The periodic reference yields −78.224 nm. Finite 180°/360° controls give −41.157/−78.646 nm with their own backgrounds. This response detects wall rotation but cannot distinguish the sign of a purely axial domain. The additional orientation proxy uses \((m_{s,\rm bg}-m_s)/2\) and an axial reference. It is not a solved tunnel-junction response.

Both local proxies use a 40 nm window with consistent cross-sectional weighting. For ten 4 nm cells, the window at index \(i\) averages \(i-5\) through \(i+4\) and lies half a cell before the associated magnetization sample. Periodic tracks wrap; finite tracks use supported interior windows. The implementation preserves constant fields and commutes with circular shifts. On the lower-current train, \(L_{\rm AMR}\) is −305.349 nm initially, −292.013 nm at pulse end and −307.414 nm after recovery. The recovered state retains separated features with orientation-proxy maximum 0.99244. This is simulated contrast, not a measured bit-error rate or electrical decoding experiment.

The single pair's excess moment is approximately \(-1.327\times10^{-16}\) A m² along the tube. An explicitly isolated-cell diagnostic has remote field scaling close to \(r^{-3}\). Two such distant aligned dipoles attract as \(U\simeq-2(\mu_0/4\pi)\mu_{\rm bit}^2/D^3\), with center separation \(D\). At the demonstration pitch, texture deformation and periodic neighbors preclude interpreting that asymptote as an optimized spacing law. Leading-charge cancellation does not imply zero dipole, negligible crosstalk or stable inter-bit distances.

%% file: sections/S10.tex
\section{Thermal scope and unresolved device questions}\label{sec:S10}
If a single activated pathway and attempt time \(\tau_0\) are meaningful, its mean escape time has the conditional Arrhenius form \(\tau\sim\tau_0\exp(\Delta E/k_BT)\), where \(\Delta E\) is the escape barrier, \(T\) is temperature and \(k_B\) is the Boltzmann constant. Holding the approximate reference barrier unchanged at 300 K would give \(\Delta E/k_BT\approx93\). The prefactor, temperature dependence, competing pathways and continuum corrections are not established, so this number is a conditional energy scale rather than a retention prediction. Survival under deterministic forcing is not a substitute for a thermal escape calculation.

The model stabilizes the pair through exchange, circumferential anisotropy and magnetostatics, without microscopic Dzyaloshinskii--Moriya interaction, a chiral-host phase or a compensation point. This makes suitable ferromagnetic energy ratios and geometry the relevant material requirements. Temperature would enter through those material parameters and through activation over competing barriers. Persistence across a temperature interval is therefore a plausible material-design question, not a result of these fixed-parameter trajectories. Local readout, prepared-pattern recovery and finite curved travel demonstrate useful ingredients; selective writing, retention statistics, optimum bit density and a complete read--write device remain distinct quantitative tasks.

%% file: sections/movies.tex
\newcounter{suppmovie}
\renewcommand{\thesuppmovie}{S\arabic{suppmovie}}
\section*{Supplementary movies}\addcontentsline{toc}{section}{Supplementary movies}
The supplementary movies show saved magnetic vectors on a smooth display support viewed with a perspective camera. The display glass is thicker than the simulated magnetic shell. Each glyph copies a nearest occupied-cell vector without projecting it onto the surface or changing its magnitude. Intermediate video frames hold a recorded magnetic state while the camera can continue moving; they do not create new magnetic dynamics. Native scalar diagnostics, rather than the visual glyph subsample, define intact intervals.

\paragraph{Movie S1.}\refstepcounter{suppmovie}\label{movie:straight}Dense straight-tube transport. One wall travels 200.81 nm over 1.93 ns at $J_s=-2\times10^{12}$ A/m$^2$ in the 14.1 nm periodic shell (3 nm cells), remaining intact at the endpoint. This is the dense saved-state replay of that thick-shell experiment, not the separate 4 nm transport series of main Fig.~\ref{fig:transport}. There are 194 saved states, displayed over 4 s at 60 frames/s. Physical time begins at current-on.

\paragraph{Movie S2.}\refstepcounter{suppmovie}\label{movie:train}Prepared-pattern forward, off and reverse motion. The 4 nm shell carries four textures encoding $1100101$ with an empty periodic guard slot. The chronological protocol is $-4\times10^{12}$ A/m$^2$ for 2 ns, zero current to 2.8 ns, then $+4\times10^{12}$ A/m$^2$ to 3.8 ns, without intervening numerical relaxation. The drive is torque-only, with no Oersted field. All four survive. There are 213 saved magnetic states, displayed over 7 s at 60 frames/s; labels identify the drive phase and continuous physical time.

\paragraph{Movie S3.}\refstepcounter{suppmovie}\label{movie:torus}Extended transport on a full torus. A 704 nm major-radius torus with a 14.1 nm shell and 3 nm cells carries physical toroidal current of cross-sectional mean $-10^{12}$ A/m$^2$. The wall remains intact through 490.065 nm (39.884°) at 9.43766 ns; the subsequent radial deformation is retained through the 9.65866 ns endpoint and labelled unresolved. The geometry is a complete torus, while the travelled path is a fraction of a circuit. The 425 saved magnetic states are displayed over 12 s at 60 frames/s, without looping or interpolating spins. The clock begins at the original current-on; the 3.53766 ns checkpoint join introduces no new relaxation.

%% file: sections/supporting_figures.tex
\begin{figure}[p]\centering\includegraphics[width=\linewidth]{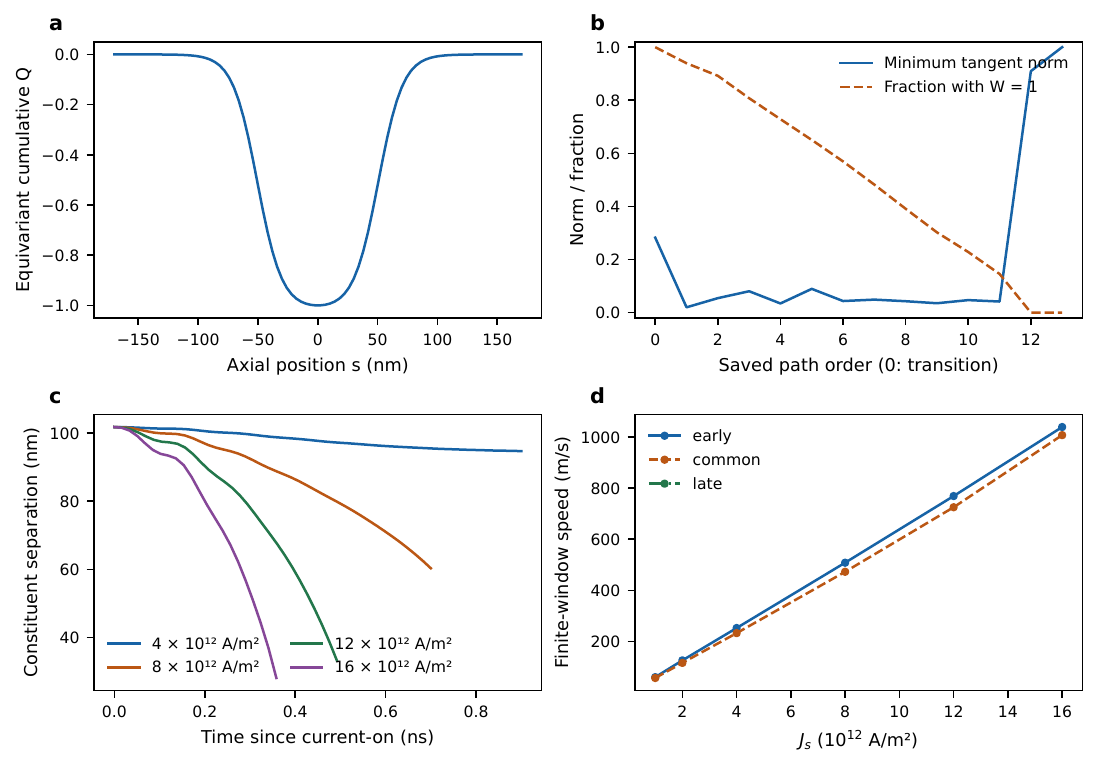}
\caption{\textbf{Charge structure, projected escape and transient transport.}
\textbf{a}, Equivariant cumulative charge $[\langle m_s(s)\rangle-\langle m_s(\mathrm{left})\rangle]/2$ of the relaxed reference; this illustrates opposite sectors and is distinct from full-field solid-angle checks. Current is zero. \textbf{b}, Minimum occupied-cell tangent norm and fraction of resolved columns with $W=1$, at the transition (index zero) and during saved energy descent. Path ordering is not time. \textbf{c}, Constituent separation for positive currents $4,8,12,16\times10^{12}$ A/m$^2$ before identity loss. \textbf{d}, Speed fits from 0.04--0.14 ns (early), 0.04--0.30 ns (common) and 0.20--0.30 ns (late) with the Oersted field included. Clocks in c/d start at current-on. Window sensitivity describes deterministic transients, not statistical confidence intervals.}
\label{fig:diagnostics}\end{figure}
\begin{figure}[p]\centering\includegraphics[width=\linewidth]{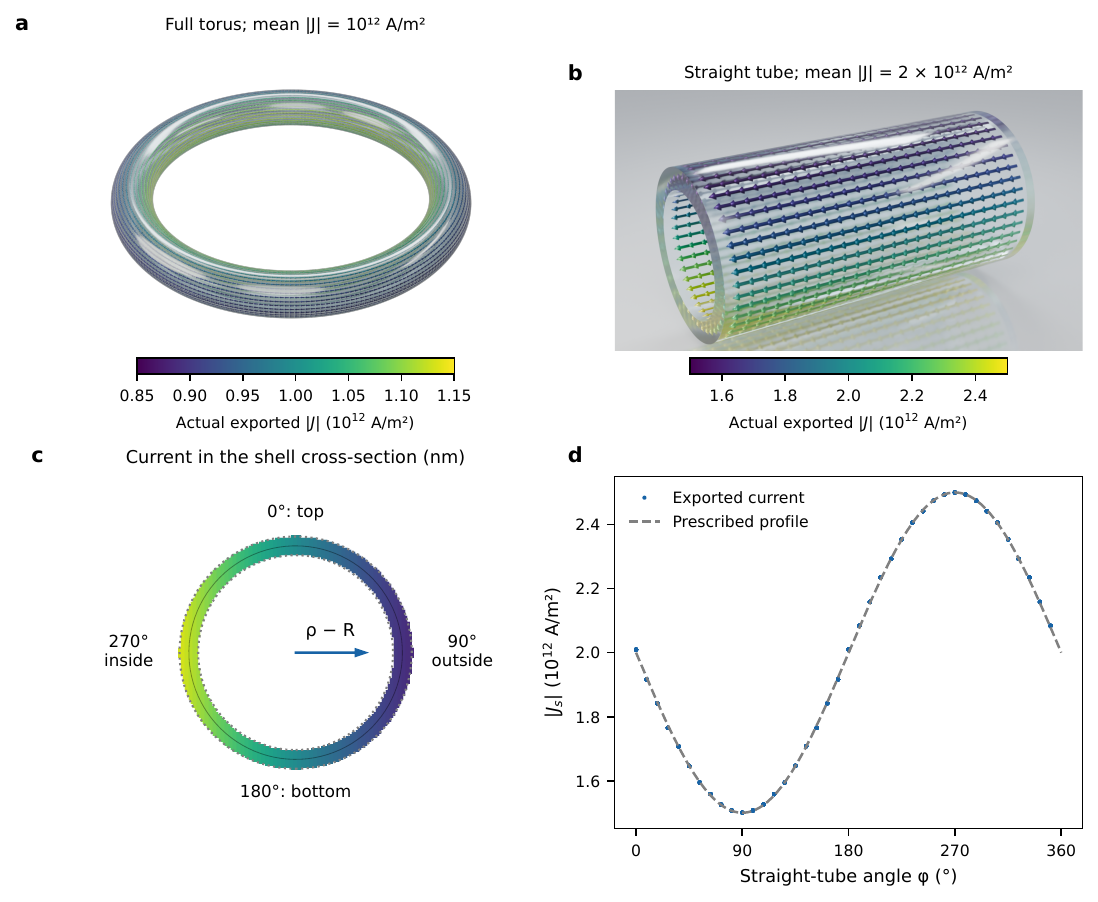}
\caption{\textbf{Actual current fields and their coordinate conventions.}
\textbf{a}, Native exported charge-current vectors on the complete torus at mean conventional current $-10^{12}$ A/m$^2$, displayed at 8,192 regular middle-surface sites. Each glyph copies the nearest occupied Cartesian-cell vector; the maximum source distance is 2.282 nm and every site has a distinct source cell. The display lattice is regular, while its vectors and magnitudes remain those of the stepped mesh. All glyphs have an 18.074 nm display length; colour, rather than length, gives $|\mathbf J|$ on the fixed scale. The middle-surface samples span 0.890--1.124 $\times10^{12}$ A/m$^2$. The glass locates the toroidal material. \textbf{b}, The exported straight-tube current with 25\% circumferential modulation. \textbf{c}, The torus magnetic-shell cross-section coloured by nearest occupied-cell current magnitude, using the same scale as a; its hollow lumen is left blank. The maximum source distance is 3.279 nm. Poloidal 0° is top, 90° outside, 180° bottom and 270° inside. \textbf{d}, Exported straight-tube samples compared with the prescribed current profile; here $\phi=\operatorname{atan2}(z,y)$. These fields are imposed charge current, not a texture-dependent spin-current reconstruction. Supplementary Fig.~\ref{fig:differential-maps} places the straight current and magnetic response on a common unrolled chart.}
\label{fig:current-fields}\label{fig:curvature-readout}\end{figure}
\begin{figure}[p]\centering\includegraphics[width=\linewidth]{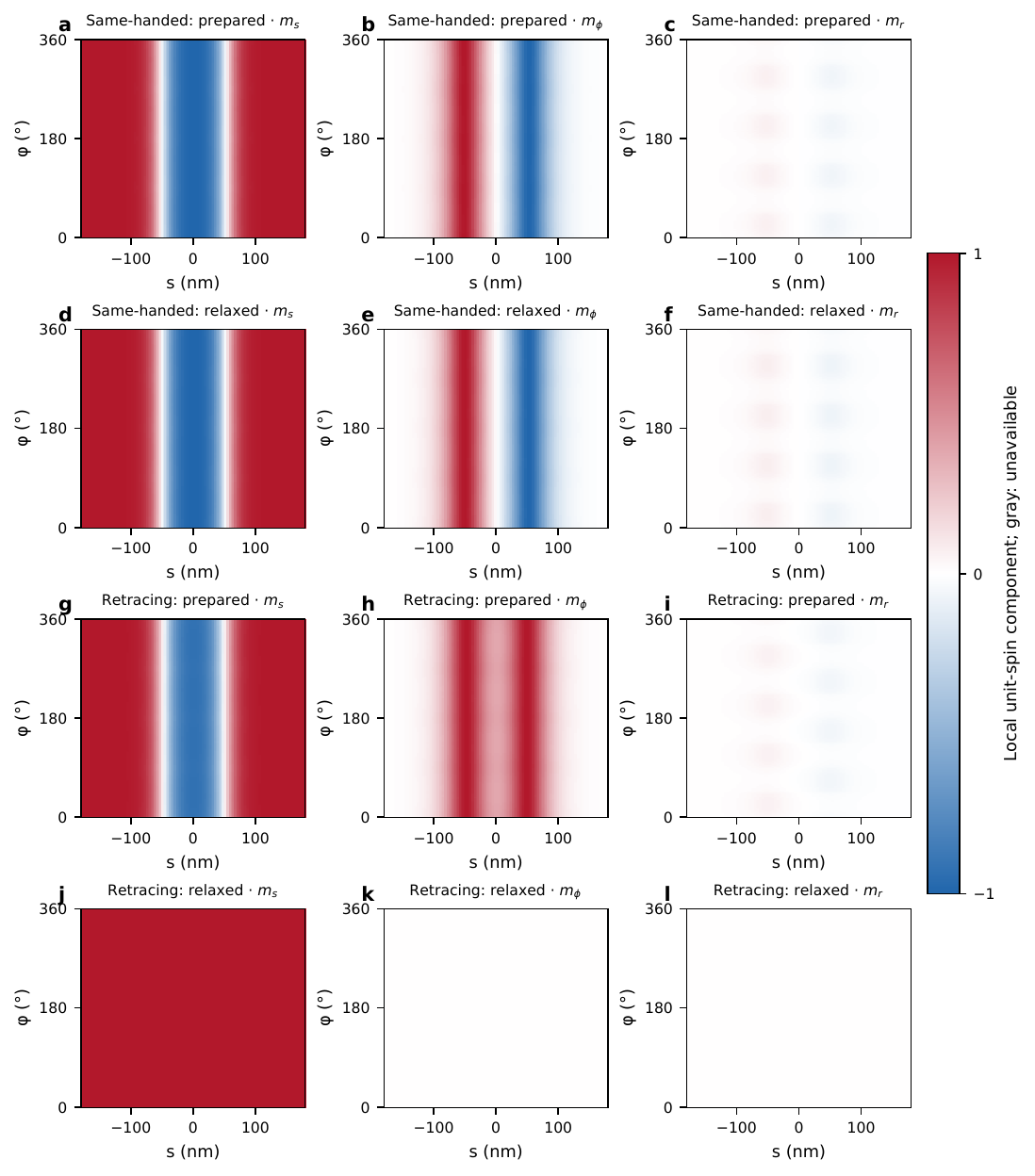}
\caption{\textbf{Same-handed and retracing textures in real space.} Each row gives the complete unrolled middle-surface components of one saved state in the 4 nm axial reference: prepared and relaxed same-handed pair, then prepared and relaxed retracing control. The three columns are $m_s,m_\phi,m_r$. All components share the full $[-1,1]$ scale, with zero white; gray denotes unavailable support. No tangent-plane constraint, axisymmetric constraint or maintaining field is applied. The retracing texture contracts to the axial background whereas the same-handed pair remains bound. The finite plotted axial window contains the complete localized texture; angles cover the complete circumference. These are preparation/relaxation endpoints, not a physical-time sequence.}
\label{fig:stability}\end{figure}
\begin{figure}[p]\centering\includegraphics[width=\linewidth]{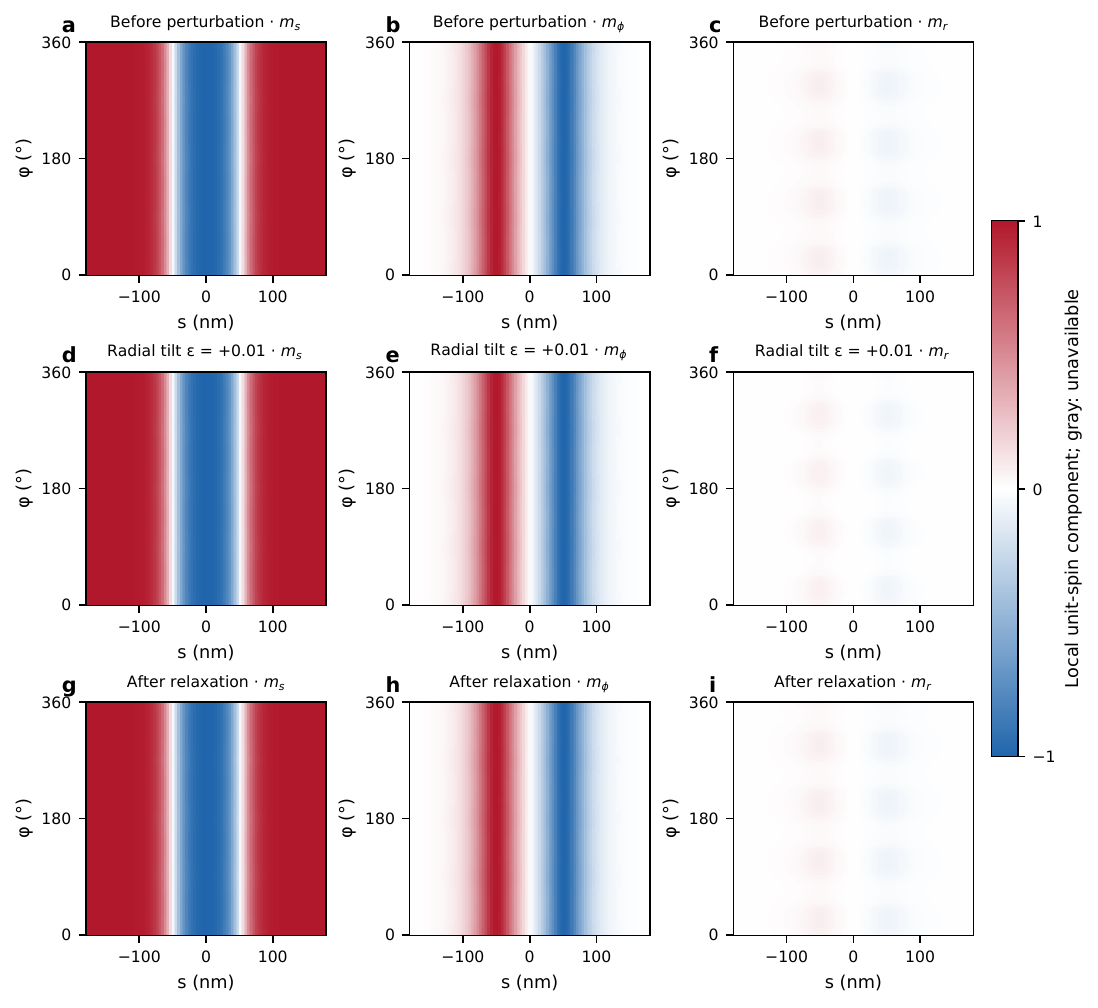}
\caption{\textbf{Radial tilt and relaxation.} Full middle-surface component maps before perturbation, immediately after the $\epsilon=+0.01$ radial tilt, and after unconstrained relaxation. Each column shares the full $[-1,1]$ scale; zero is white and unavailable support gray. The RMS spin-angle difference from the common unperturbed reference in Table~\ref{tab:perturbations} decreases from 0.2631° to 0.0153° after registering only common axial translation. The maximum imposed spin change is approximately 0.01. Table~\ref{tab:perturbations} quantifies all ten signs/modes. Current and maintained perturbation fields are zero. Numerical relaxation is not physical-time dynamics.}
\label{fig:perturbation-maps}\end{figure}
\begin{figure}[p]\centering\includegraphics[width=\linewidth]{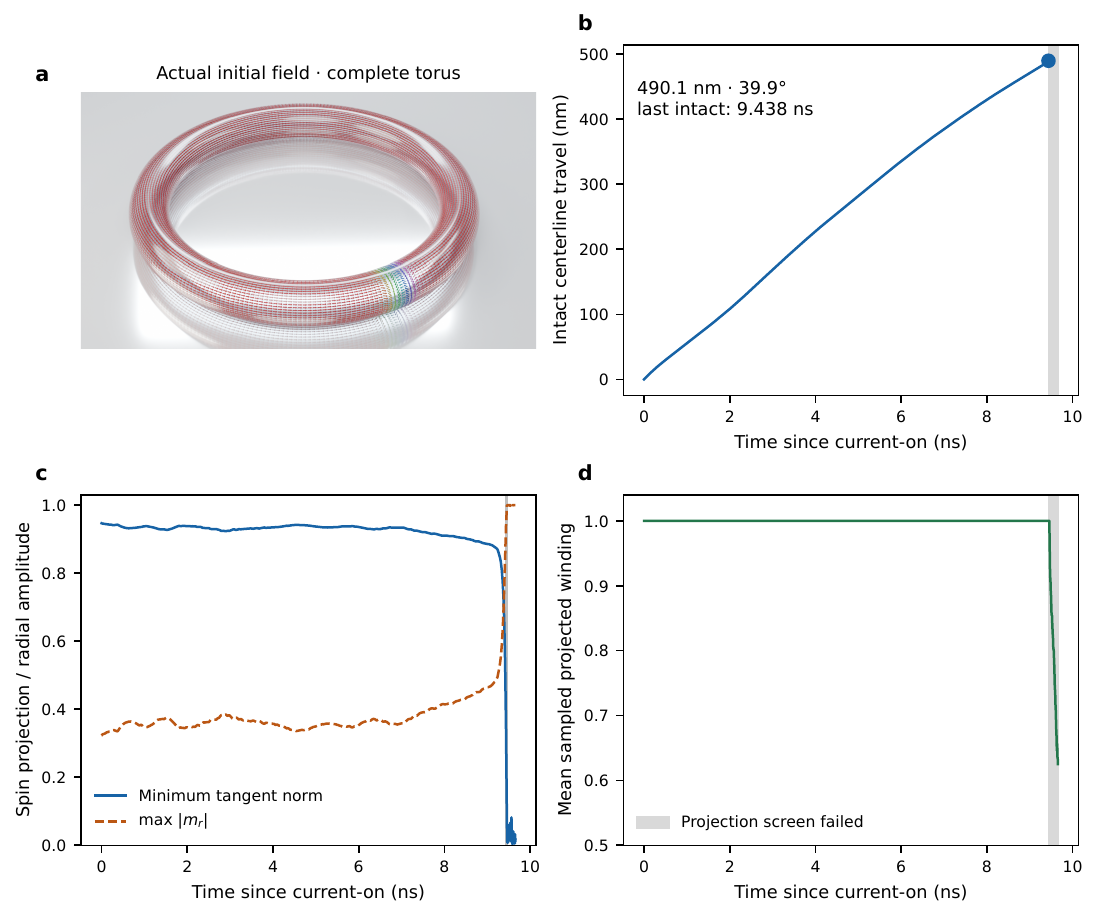}
\caption{\textbf{Extended intact transport on a complete closed torus.}
\textbf{a}, Actual prepared magnetization at drive-on. The 704 nm major radius and 14.1 nm magnetic shell close into a full torus; there are no magnetic end faces or Cartesian periodic boundaries. \textbf{b}, Intact centerline travel reaches 490.065 nm (39.884°) by 9.43766 ns under mean conventional current $-10^{12}$ A/m$^2$, with the physical toroidal profile of Fig.~\ref{fig:current-fields}a. \textbf{c}, Minimum tangent norm and maximum radial amplitude, including the late deformation. The narrow band brackets the first failed screen between 9.43766 and 9.45766 ns. \textbf{d}, Mean sampled winding stays one through that first projection event and becomes nonuniform later; the gray interval is unsuitable for interpreting an integer winding as a protected identity. Each clock begins at the original current-on, with exact-checkpoint continuation and no intervening preparation. The final 9.65866 ns endpoint is an unresolved deformation, not an established terminal annihilation or identified numerical artifact. Neither a completed revolution nor an exact radius/current threshold is inferred. Supplementary Movie~\ref{movie:torus} preserves the full recorded sequence.}
\label{fig:curvature}\end{figure}
\begin{figure}[p]\centering\includegraphics[width=\linewidth]{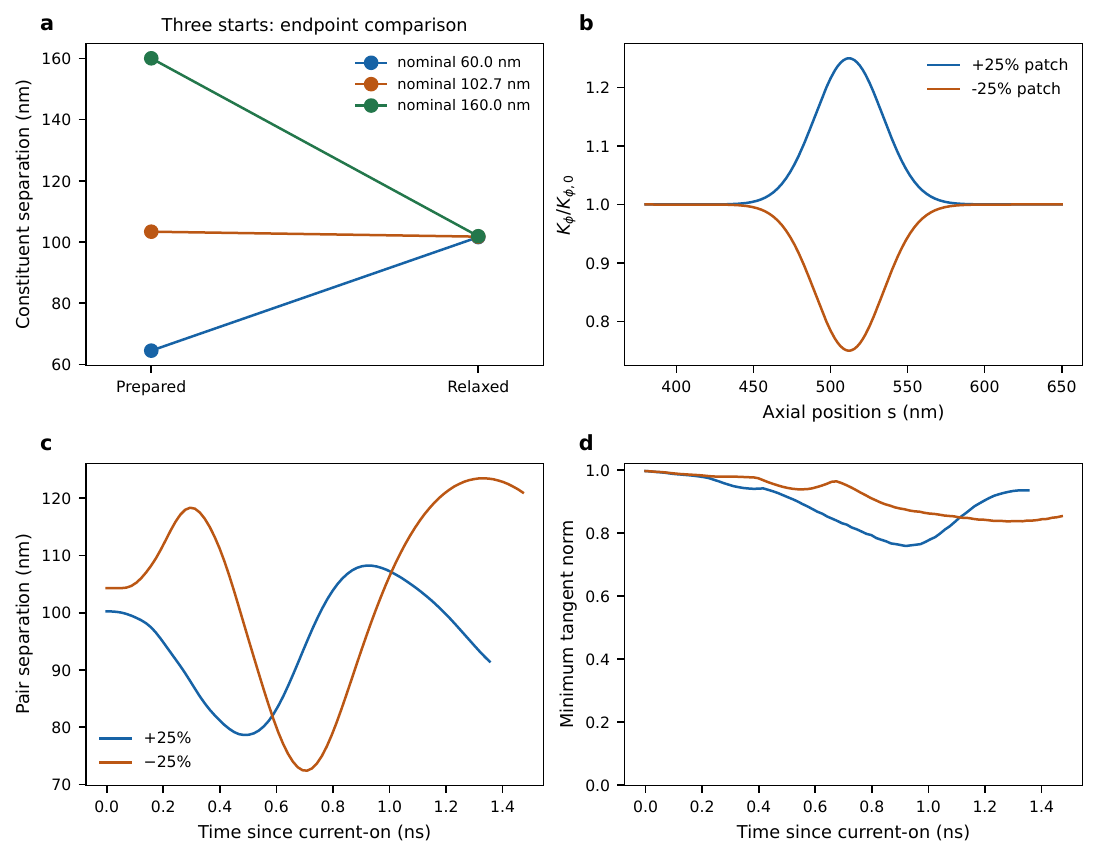}
\caption{\textbf{Restoration from different separations and traversal of local anisotropy patches.}
\textbf{a}, Measured starting and relaxed separations for three nominal initialization parameters. Lines connect endpoints of numerical relaxation; they are not kinetic trajectories. The short overlapping profile's measured zero-crossing separation is 64.49 nm for a nominal 60 nm input. All approach approximately 102 nm. \textbf{b}, Prescribed spatial patch profiles of circumferential anisotropy, Gaussian width parameter 22 nm and amplitude $\pm25\%$, centered at $s=512$ nm. \textbf{c,d}, Measured separation and minimum tangent norm during traversal at $J_s=4\times10^{12}$ A/m$^2$, with physical time measured from current-on. The driven phase follows preparation and 20 ps at zero current. Both trajectories retain their wall through approximately 240 nm of displacement; their different terminal times reflect the recorded observation target. These use the 4 nm shell, full vector dynamics and the ideal annular Oersted field.}
\label{fig:robustness}\end{figure}
\begin{figure}[p]\centering\includegraphics[width=\linewidth]{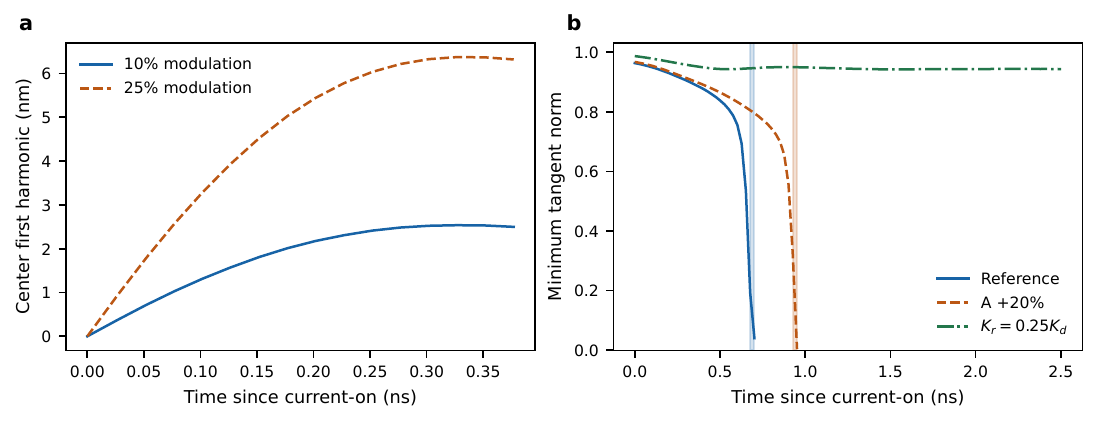}
\caption{\textbf{Differential forcing and an independently prepared radial-confinement control.}
\textbf{a}, First-harmonic center distortion of straight-tube walls under 10\% and 25\% modulation of mean current $-2\times10^{12}$ A/m$^2$. The fitted initial rates scale in a ratio of 2.5003 (Section~\ref{sec:S6}). \textbf{b}, Minimum tangent norm for 25\% modulation with the reference energy, 20\% larger exchange, or an added radial penalty $K_r=0.25K_d=16.25$ kJ/m$^3$, where $K_d=\mu_0M_s^2/2$. Shaded intervals bracket first projection-screen events. The radial-penalty case remains intact through 201.4 nm. All clocks start at current-on. These are separately prepared 14.1 nm straight shells with full demagnetization; no Oersted field is applied. The intervention is distinct from circumferential anisotropy and has not been tested as a torus barrier or torus transport control.}
\label{fig:differential-control}\end{figure}
\begin{figure}[p]\centering\includegraphics[width=\linewidth]{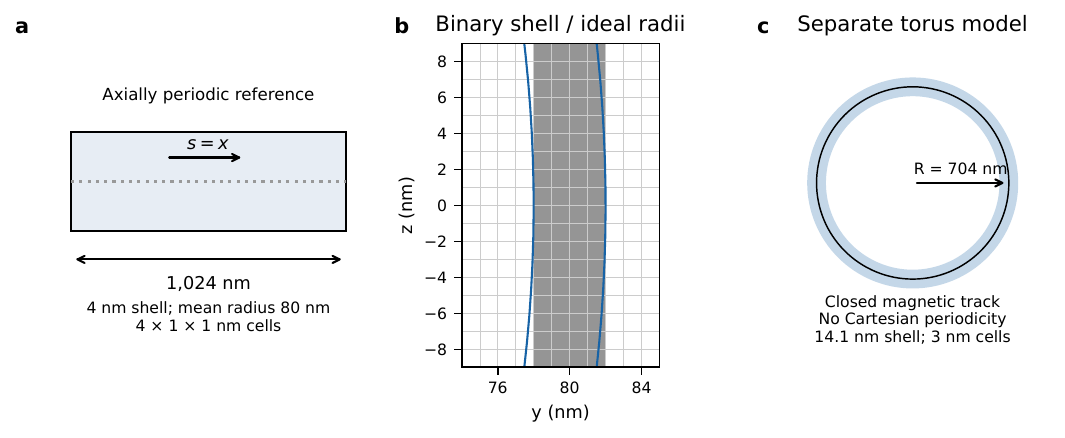}
\caption{\textbf{Geometry, discretization and boundary conventions.} Explanatory diagrams of the model families, not additional magnetic data. \textbf{a}, Straight reference with axial coordinate $s=x$, 1,024 nm periodic length, mean radius 80 nm and 4 nm magnetic shell. \textbf{b}, Magnified cross-sectional binary-cell construction at the unshifted placement, showing the 1 nm transverse cells against the ideal inner/outer radii 78/82 nm. This distinguishes the smooth display support from the Cartesian magnetic occupancy. \textbf{c}, Complete torus of major radius 704 nm, with a separately prepared 14.1 nm shell and 3 nm cubic mesh. There are no physical end faces or Cartesian periodic boundaries. Extended Methods gives the pattern and finite-ended variants.}
\label{fig:model-geometries}\end{figure}

\begin{figure}[htbp]
\centering\includegraphics[width=\linewidth]{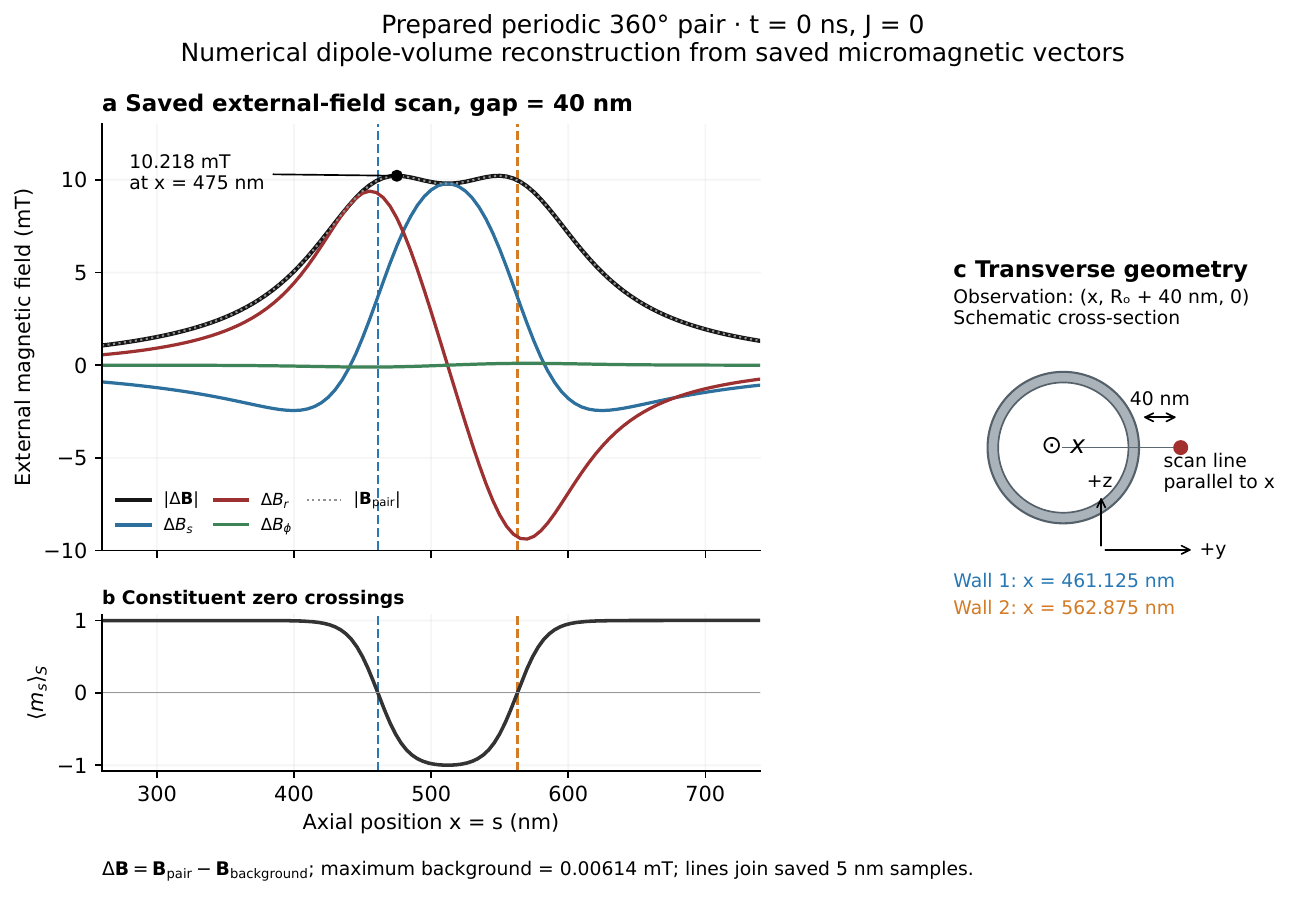}
\caption{\textbf{External stray-field contrast of the prepared pair.}
\textbf{a}, Numerical dipole-volume reconstruction on the line $(x,R_o+40\,\mathrm{nm},0)$, using the saved magnetic vectors, occupied cells and four periodic images in each axial direction. Solid curves give the magnitude and components of $\Delta\mathbf B=\mathbf B_{\rm pair}-\mathbf B_{\rm background}$; here $B_s=B_x$, $B_r=B_y$ and $B_\phi=B_z$. The sampled maximum is 10.218 mT at $x=475$ nm. The dotted absolute pair-field magnitude almost coincides (maximum 10.214 mT); the maximum background magnitude is 0.00614 mT. Curves join 5 nm-spaced samples. \textbf{b}, The same state's mean axial magnetization locates the two constituent centers at 461.125 and 562.875 nm, marked in matching colours in a. \textbf{c}, Schematic transverse geometry and the 40 nm gap from the outer surface. The state is post-preparation at $t=0$ and $J=0$. The external field is reconstructed numerically from the saved magnetization. Source-quadrature checks are in Supplementary Section~\ref{sec:S9}; no continuous-space global maximum or detector response is inferred.}
\label{fig:point-readout}
\end{figure}
\begin{figure}[htbp]
\centering\includegraphics[width=\linewidth]{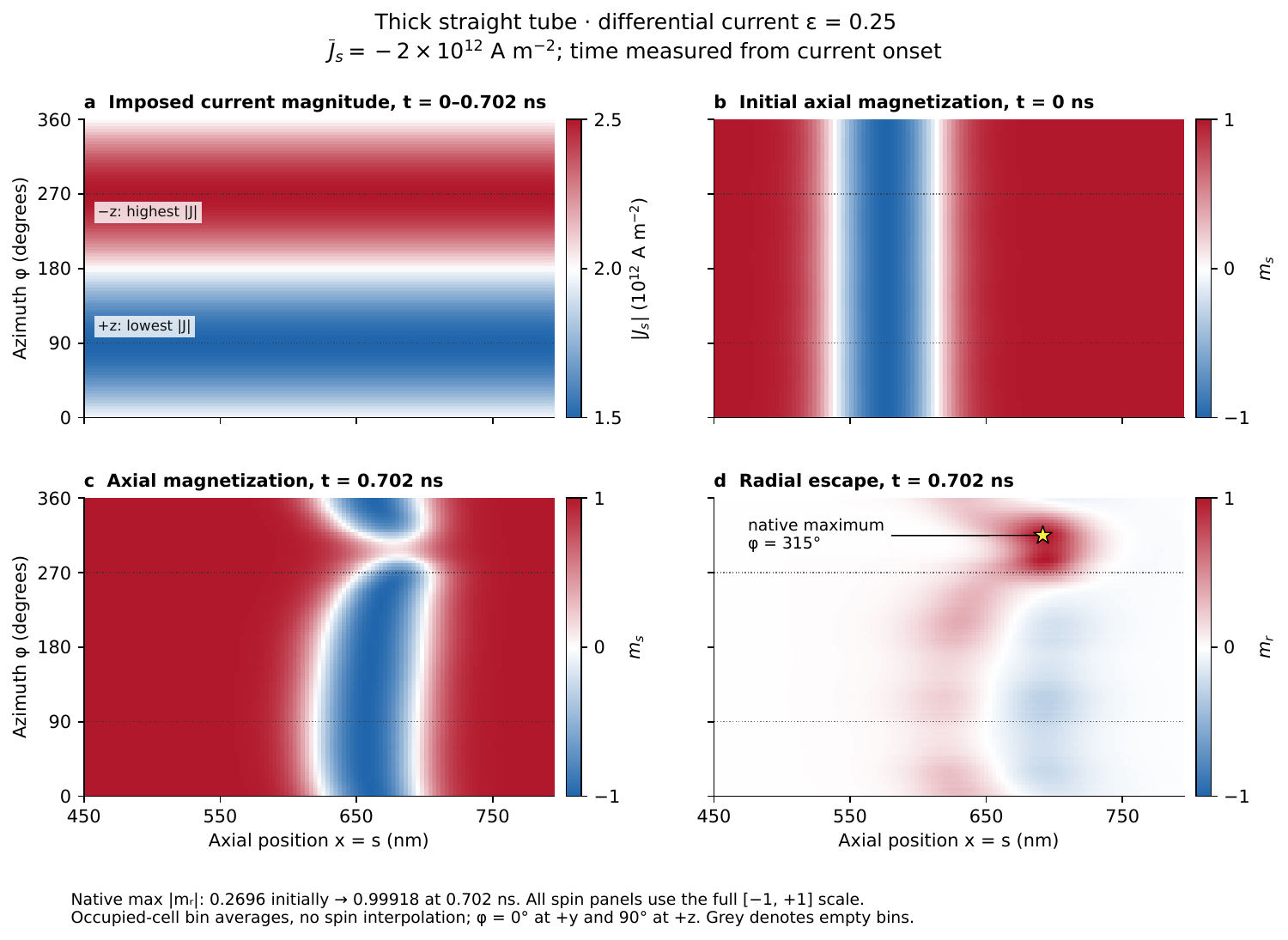}
\caption{\textbf{Nonuniform axial current and the associated radial escape channel.}
All panels share axial position and $\phi=\operatorname{atan2}(z,y)$ in the thick straight-tube control. \textbf{a}, Exported $|J_s|$ for mean signed current $-2\times10^{12}$ A/m$^2$ and 25\% modulation. Current is uniform along $x$ and remains on. Its magnitude is smallest at 90° and largest at 270°. \textbf{b}, Initial $m_s$ at current-on. \textbf{c,d}, $m_s$ and $m_r$ at 0.702 ns, the first saved sample failing the projection screen after a passing sample at 0.677 ns. The star projects the full-volume maximum $|m_r|=0.99918$ at $x=691.5$ nm, $\phi=315$° onto the chart. It lies on the higher-current half, away from the current maximum. Maps average occupied-cell components in 3 nm $\times$ 3.75° bins over cells at or above the 55th radius percentile; this outer slice contains 45.455\% of cells including ties. The starred cell itself is at an inner radius, whereas the displayed outer-bin maximum is 0.99578. Spin scales remain $[-1,1]$, with no vector normalization or radial amplification; unsupported bins would be gray. This nearly radial state does not establish a uniquely resolved nucleation time or completed annihilation. The shell radii are 72.929/87.071 nm, axial period 1152 nm and mesh 3 nm. Dynamics retain full demagnetization with $\alpha=0.1$, $\beta=0.05$, $P=0.4$ and no Oersted field.}
\label{fig:differential-maps}
\end{figure}

\clearpage
\begin{figure}[p]
\centering\includegraphics[width=\linewidth,height=.67\textheight,keepaspectratio]{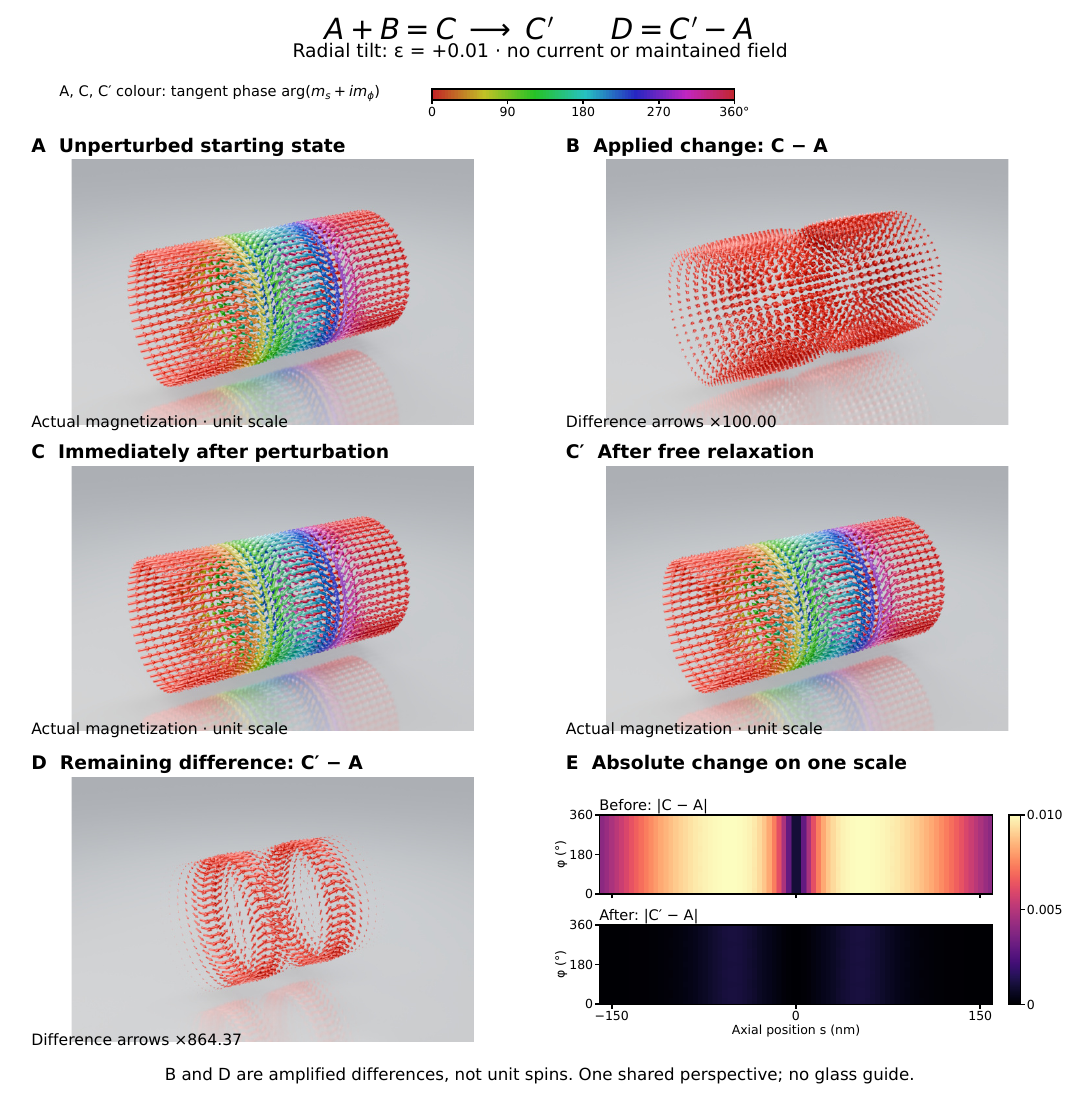}
\caption{\textbf{The applied perturbation and the remaining difference, shown separately from the spin texture.}
\textbf{A}, Exact unperturbed starting state. \textbf{B}, Applied spin-vector change $C-A$. \textbf{C}, Saved state immediately after the positive radial tilt, $\epsilon=+0.01$. \textbf{C$'$}, Saved endpoint after unconstrained relaxation. \textbf{D}, Remaining difference $C'-A$. \textbf{E}, Middle-surface magnitudes of both differences on the same absolute scale. A, C and C$'$ use identical unit-magnetization arrow scales and colours for tangent phase. B and D use one colour and global amplification factors 100.00 and 864.37; all glyph dimensions scale with vector magnitude, preserving relative amplitudes within each field. The maximum differences over all occupied cells are 0.00999963 and 0.00115692; the corresponding RMS spin angles from A are 0.26131° and 0.01832°. Difference arrows are not unit spins. The identity $A+B=C$ refers to the unamplified vectors. No translation is removed; D includes continued relaxation of the starting state and is not a numerical-error estimate. The rendered crop spans $s=-158$ to 154 nm of the 1024 nm axial period. Full Cartesian vectors are copied from the nearest occupied mesh cells onto a smooth 80 nm-radius display surface, with maximum sampling distance 0.7072 nm; no interpolation or tangent projection is used. All panels share the $4\times1\times1$ nm mesh and 78/82 nm inner/outer radii. Current and applied fields are zero. Numerical relaxation is not physical-time dynamics.}
\label{fig:perturbation-vectors}
\end{figure}

\clearpage
\begin{figure}[p]
\centering\includegraphics[width=\linewidth,height=.73\textheight,keepaspectratio]{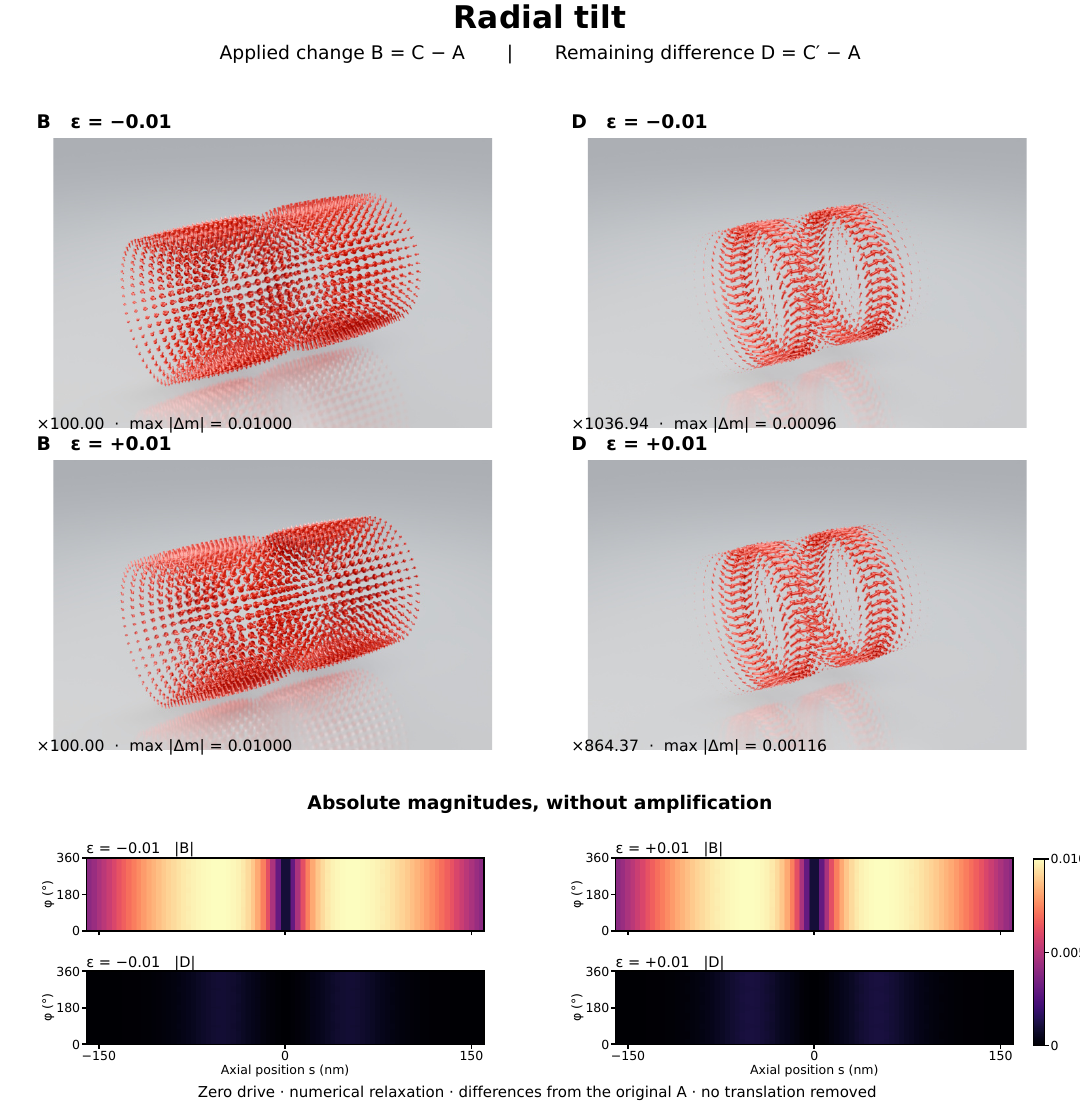}
\caption{\textbf{Radial tilt: applied and remaining spin-vector differences.} The upper render row shows $\epsilon=-0.01$ and the lower row $\epsilon=+0.01$. The left column shows $B=C-A$; the right shows $D=C'-A$. The maps below use one absolute magnitude scale, $[0,0.01]$, for both signs and stages. The occupied-volume RMS angles from the exact original starting state $A$ decrease from 0.26131° to 0.01528° for the negative sign, and from 0.26131° to 0.01832° for the positive sign. Arrow amplification is labelled in every render; it preserves directions and relative amplitudes within each field. The reference, full-vector nearest-cell sampling and zero-drive protocol are those of Fig.~\ref{fig:perturbation-vectors}. No translation is removed, so $D$ includes continued relaxation of $A$ and is not a numerical-error estimate.}
\label{fig:perturbation-radial}
\end{figure}

\clearpage
\begin{figure}[p]
\centering\includegraphics[width=\linewidth,height=.73\textheight,keepaspectratio]{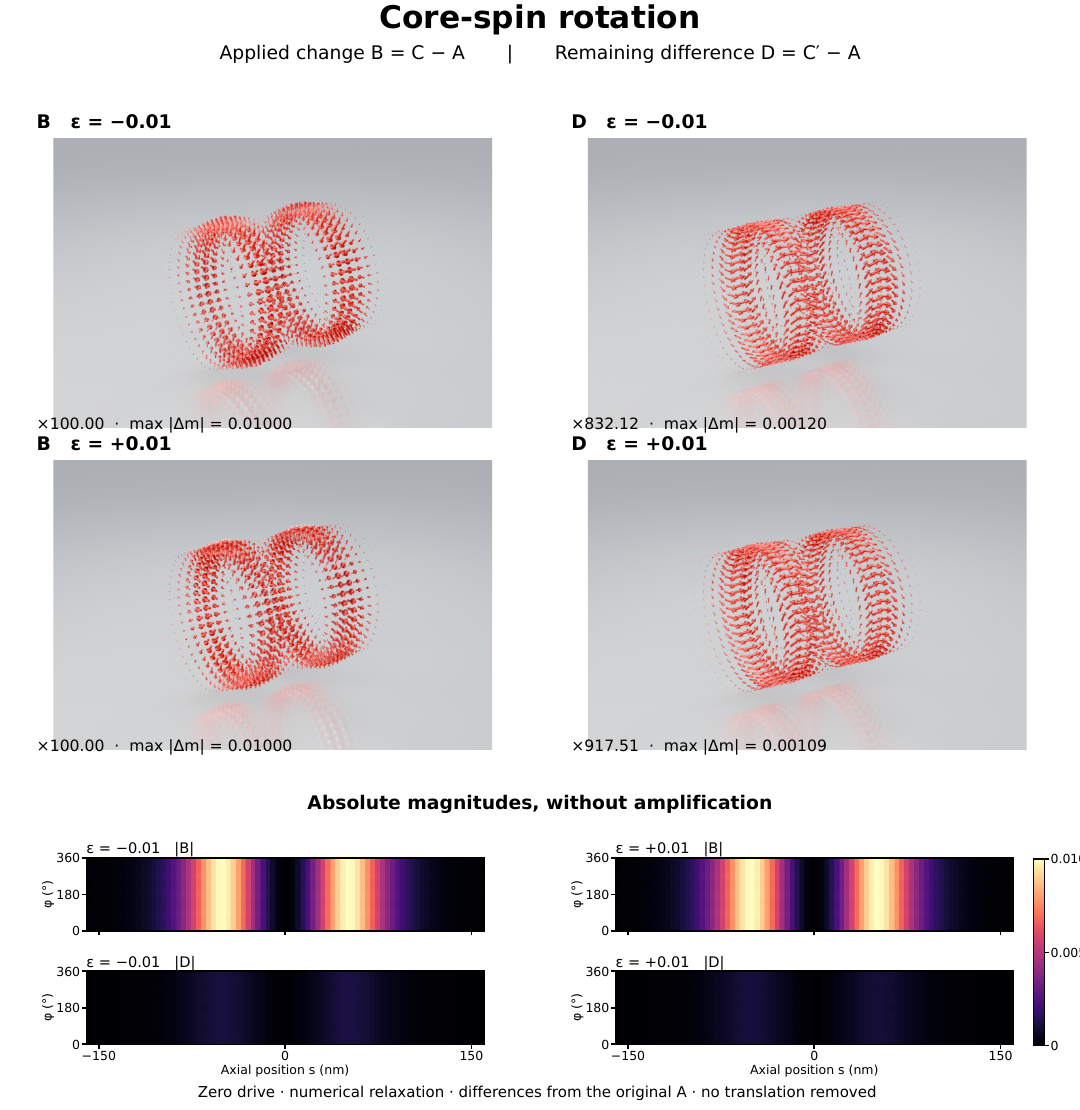}
\caption{\textbf{Core-spin rotation: applied and remaining spin-vector differences.} The upper render row shows $\epsilon=-0.01$ and the lower row $\epsilon=+0.01$. The left column shows $B=C-A$; the right shows $D=C'-A$. The maps below use one absolute magnitude scale, $[0,0.01]$, for both signs and stages. The occupied-volume RMS angles from the exact original starting state $A$ decrease from 0.15413° to 0.01855° for the negative sign, and from 0.15413° to 0.01678° for the positive sign. Arrow amplification is labelled in every render; it preserves directions and relative amplitudes within each field. The reference, full-vector nearest-cell sampling and zero-drive protocol are those of Fig.~\ref{fig:perturbation-vectors}. No translation is removed, so $D$ includes continued relaxation of $A$ and is not a numerical-error estimate.}
\label{fig:perturbation-helicity}
\end{figure}

\clearpage
\begin{figure}[p]
\centering\includegraphics[width=\linewidth,height=.73\textheight,keepaspectratio]{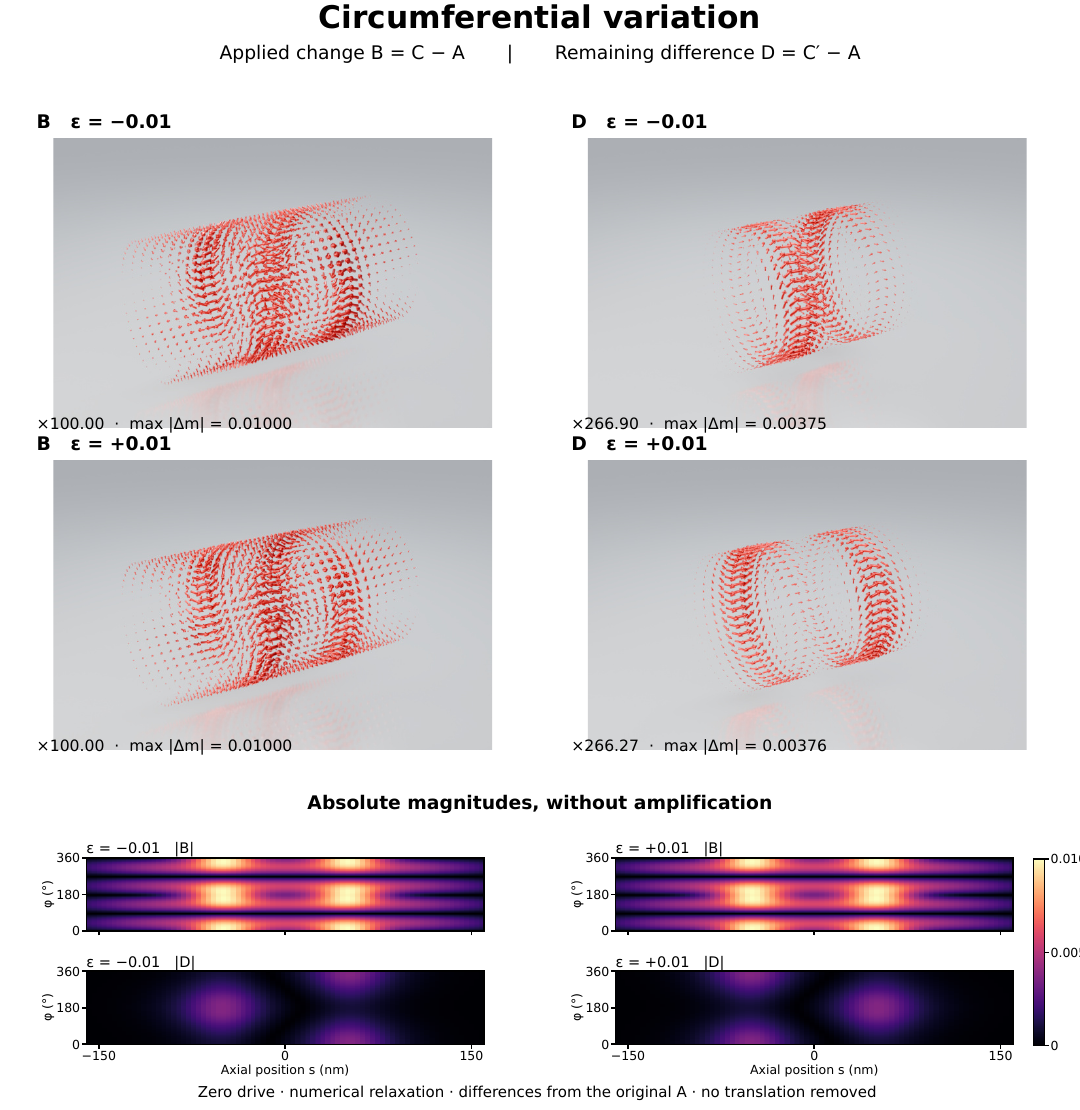}
\caption{\textbf{Circumferential variation: applied and remaining spin-vector differences.} The upper render row shows $\epsilon=-0.01$ and the lower row $\epsilon=+0.01$. The left column shows $B=C-A$; the right shows $D=C'-A$. The maps below use one absolute magnitude scale, $[0,0.01]$, for both signs and stages. The occupied-volume RMS angles from the exact original starting state $A$ decrease from 0.15588° to 0.04151° for the negative sign, and from 0.15588° to 0.04150° for the positive sign. Arrow amplification is labelled in every render; it preserves directions and relative amplitudes within each field. The reference, full-vector nearest-cell sampling and zero-drive protocol are those of Fig.~\ref{fig:perturbation-vectors}. No translation is removed, so $D$ includes continued relaxation of $A$ and is not a numerical-error estimate.}
\label{fig:perturbation-azimuthal}
\end{figure}

\clearpage
\begin{figure}[p]
\centering\includegraphics[width=\linewidth,height=.73\textheight,keepaspectratio]{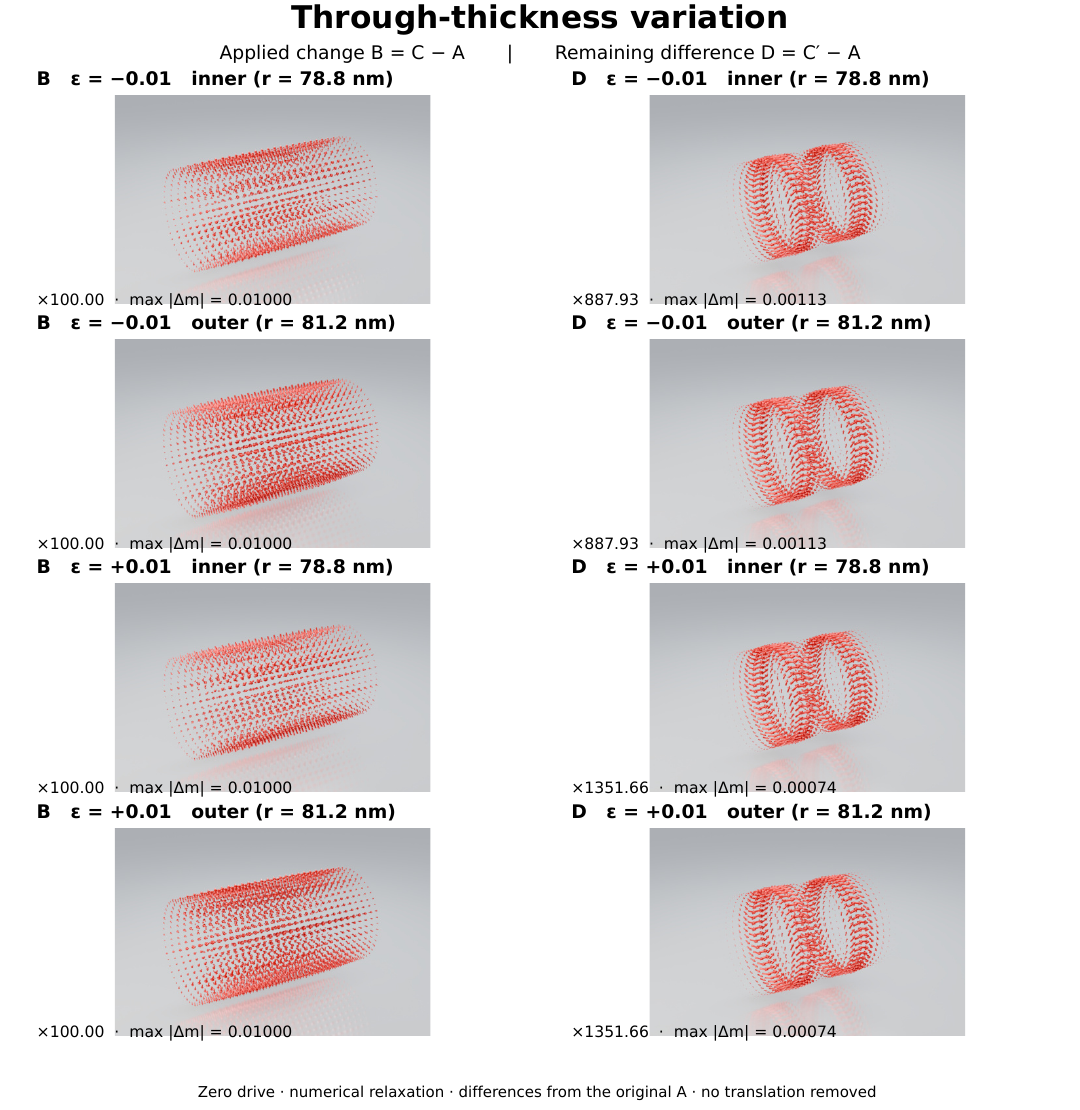}
\caption{\textbf{Through-thickness variation: applied and remaining spin-vector differences.} Rows show $\epsilon=-0.01$ at the inner and outer sample surfaces, followed by $\epsilon=+0.01$ at the same surfaces. The left column shows $B=C-A$; the right shows $D=C'-A$. The radii are 78.8 and 81.2 nm; opposite through-thickness changes are resolved rather than sampled only at the middle surface. The occupied-volume RMS angles from the exact original starting state $A$ decrease from 0.16199° to 0.01784° for the negative sign, and from 0.16199° to 0.01172° for the positive sign. Arrow amplification is labelled in every render; it preserves directions and relative amplitudes within each field. Inner and outer views share the same amplification for each field. The reference, full-vector nearest-cell sampling and zero-drive protocol are those of Fig.~\ref{fig:perturbation-vectors}. No translation is removed, so $D$ includes continued relaxation of $A$ and is not a numerical-error estimate.}
\label{fig:perturbation-thickness}
\end{figure}

\clearpage
\begin{figure}[p]
\centering\includegraphics[width=\linewidth,height=.73\textheight,keepaspectratio]{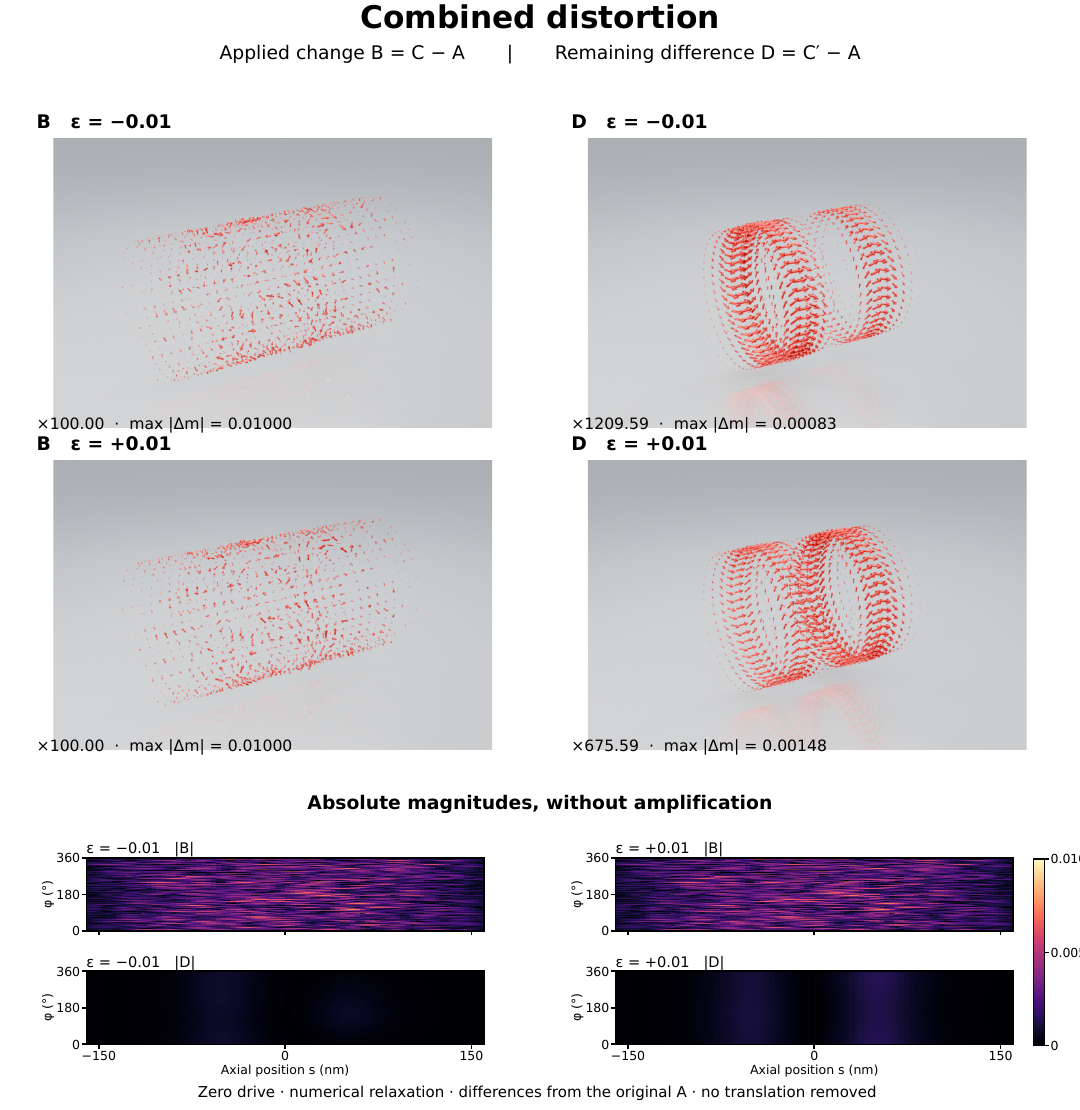}
\caption{\textbf{Combined distortion: applied and remaining spin-vector differences.} The upper render row shows $\epsilon=-0.01$ and the lower row $\epsilon=+0.01$. The left column shows $B=C-A$; the right shows $D=C'-A$. The maps below use one absolute magnitude scale, $[0,0.01]$, for both signs and stages. The occupied-volume RMS angles from the exact original starting state $A$ decrease from 0.08533° to 0.01044° for the negative sign, and from 0.08533° to 0.01977° for the positive sign. Arrow amplification is labelled in every render; it preserves directions and relative amplitudes within each field. The reference, full-vector nearest-cell sampling and zero-drive protocol are those of Fig.~\ref{fig:perturbation-vectors}. No translation is removed, so $D$ includes continued relaxation of $A$ and is not a numerical-error estimate.}
\label{fig:perturbation-mixed}
\end{figure}